\documentclass{ar-astro2e} 

\usepackage{graphicx, amssymb, multirow}
\usepackage{ARAstroBib}  \jname{Annu. Rev. Astron. Astrophys.}
\usepackage{color}

\begin{document} 

\input epsf.def 

\input psfig.sty

\jyear{2013} \jvol{51} \ARinfo{1056-8700/97/0610-00}

\title{Stellar Multiplicity}

\markboth{Gaspard Duch\^ene \& Adam Kraus}{Stellar Multiplicity}

\author{Gaspard Duch\^ene$^{1,2}$ \& Adam Kraus$^{3}$ \affiliation{$^1$University of California Berkeley, Hearst Field Annex B-20, Berkeley CA 94720-3411 USA\\ $^2$UJF-Grenoble 1/ CNRS-INSU, Institut de Plan\'etologie et
d'Astrophysique de Grenoble \\
(IPAG) UMR 5274, BP 53, 38041 Grenoble Cedex 9, France \\ $^3$Harvard-Smithsonian Center for Astrophysics, 60 Garden Street, Cambridge, MA 02138, USA}}


\begin{abstract} 
Stellar multiplicity is an ubiquitous outcome of the star formation process. Characterizing the frequency and main characteristics of multiple systems and their dependencies on primary mass and environment is therefore a powerful tool to probe this process. While early attempts were fraught with selection biases and limited completeness, instrumentation breakthroughs in the last two decades now enable robust analyses. In this review, we summarize our current empirical knowledge of stellar multiplicity for Main Sequence stars and brown dwarfs, as well as among populations of Pre-Main Sequence stars and embedded protostars. Clear trends as a function of both primary mass and stellar evolutionary stage are identified that will serve as a comparison basis for numerical and analytical models of star formation.
\end{abstract} 

\maketitle

\section{INTRODUCTION}
\label{sec:intro}

\indent \indent The existence of binary and multiple systems was recognized soon after the birth of modern astronomy in the 17$^{\rm th}$ century. \cite{kuiper35} first argued that empirically determining the multiplicity frequency and distribution of key orbital parameters would prove to be highly valuable from a theoretical standpoint. The dependency of the multiplicity frequency and their associated orbital parameters on primary mass should contain the imprint of the physical processes at play throughout the lifetime of stellar populations. In particular, if dynamical interactions (within multiple systems, in stellar clusters and in the Galactic tidal field) can be properly accounted for, statistical properties of multiple systems can serve as a posteriori tests for star formation theories. For instance, whether star formation is characterized by universal multiplicity distributions (akin to the apparently-universal Initial Mass Function, IMF, in the present day Universe) is an open question that has important bearing on star formation theory. 

The frequency and properties of multiple systems have been discussed in multiple entries in the Annual Review: \cite{vandekamp75} discussed astrometric binaries, while \cite{abt83} and \cite{mathieu94} presented statistical analyses of multiplicity among Main Sequence (MS) field stars in and Pre-Main Sequence (PMS) stars, respectively. Knowledge of stellar multiplicity has greatly expanded since these seminal reviews, due to improvements in instrumentation and larger, more uniform samples. In addition, new stellar populations have been probed, such as high-mass stars \citep{zinnecker07} and very low-mass (VLM) stars and brown dwarfs \citep[BDs, see][]{luhman12}, allowing for a more complete picture to be drawn. On the theoretical side, prompt fragmentation during the gravitational collapse of a prestellar core appears to be the favored formation mechanism although no single process appears to account for all observed multiplicity properties \citep{tohline02}.

This review summarizes our current empirical knowledge of the frequency and principal properties (such as the distributions of orbital periods, mass ratios, and eccentricities) of multiple systems as a function of stellar mass and age. Multiple systems are thought to play central roles in numerous aspects of post-MS evolution, such as the blue straggler phenomenon, X-ray binaries, Type\,Ia supernovae, and the shaping of bipolar planetary nebulae. However, we restrict the scope of this review to MS and less evolved objects, where unbiased uniform surveys have targeted larger samples. 
 
\section{CHARACTERIZING MULTIPLICITY AND SURVEY METHODOLOGY}
\label{sec:methods}

\subsection{Key multiplicity properties}

\indent \indent The frequency of multiple systems and the companion frequency (average number of companion per target, which can exceed 100\%), $MF$ and $CF$ respectively, are the  most immediately quantifiable aspect of multiplicity. The difference between these two quantities hinges on the frequency of high-order multiple systems. Unfortunately, these two quantities are frequently confused in the literature, leading to ambiguous interpretations. Beyond these global quantities, and although the diversity of multiple systems is extreme, their statistical properties can be summarized in a handful of distributions, which we briefly introduce here. While correlations between these distributions could be present, multi-dimensional analyses require very large samples of binaries, which remain out of reach in most cases.

The orbital period, $P$, of a binary system can be readily estimated for spectroscopic binaries (SBs). For visual binaries (VBs), the orbital semi-major axis, $a$, can be estimated from the projected separation, $\rho$, on a statistical basis \citep[e.g.,][]{brandeker06}. A power law parametrization, $f(P) \propto P^{\alpha}$, is commonly used, with the particular case $\alpha = -1$ \citep[a logarithmically-flat distribution known as ``\"Opik's law'',][]{opik24} being frequently favored. An alternative log-normal representation, parametrized by $\overline{P}$ and $\sigma_{\log P}$, is also widely used, although it differs only marginally from \"Opik's law for large values of $\sigma_{\log P}$ \citep{heacox96}. Distinguishing empirically between these two parametrizations would have fundamental implications on the binary formation process, as \"Opik's law suggests a scale-free process while a log-normal function implies a preferred spatial scale for companion formation. However, dynamical evolution throughout a system's lifetime can significantly alter its period, complicating the interpretation.

The mass ratio, $q = M_{\rm sec} / M_{\rm prim} \leq 1$, can generally be derived from the observed flux ratio, although only a model-dependent estimate can be obtained for PMS and substellar objects. A power law distribution, $f(q) \propto q^\gamma$, is frequently used to characterize the distribution of mass ratios and we use this formalism in this review to compare different populations. When no robust estimate of $\gamma$ has been proposed in the literature, we computed it from available data. The orbital eccentricity is an important tool to probe the dynamical evolution of a system: short period binaries can circularize as a result of tidal dissipation \citep{koch81}, while eccentricity can be pumped in three-body internal (Kozai) and external interactions. The expected distribution for a dynamically relaxed (``thermal'') population is $f(e)=2e$ \citep{ambartsumian37}. Because eccentricity can only be reliably estimated for binaries whose orbit is short enough to be fully mapped, only a subset of all systems can be included in the analysis of the eccentricity distribution. We limit our discuss of this quantity to Section\,\ref{subsec:ecc}. Finally, the other orbital parameters define the orientation of the orbital plane relative to our viewing geometry and are therefore not as astrophysically compelling. The orientation of orbital planes is essentially random with respect to our line of sight and to the stellar rotation spin vector \citep[e.g.,][]{popovic04, hale94}, while multiple systems only have a modest preference for internal coplanarity \citep[e.g.,][]{sterzik02}. We will not discuss these quantities further in this review.

\subsection{General methodology}

\indent \indent An optimal multiplicity survey should 1) rely on a complete, uniform search of a large, volume-limited sample and 2) make use of a variety of observing methods, as each of them is tailored to a certain type of companions. Magnitude-limited surveys, which are very common, are affected by an inevitable Malmquist-like bias, but this effect can be corrected statistically. Still, such surveys can be affected by more subtle selections biases, such as preferences for less extincted and/or more massive (brighter) targets. A sample size of at least 100 targets is typically required to measure multiplicity frequencies with a precision of $\pm 5 \%$. Studies of orbital parameters (e.g., $P$ or $q$) require a similar number of companions, hence a sample that is several times larger. 

Whenever possible, the comparison between two independent surveys should be performed directly over a common section of the parameter space, and binomial statistics should be preferred \citep[see][]{brandeker06}. However, most multiplicity surveys are incomplete and it is normal practice to estimate a completeness correction to take ``missed companions'' into account. Bayesian inference techniques can also be used to optimally exploit incomplete data while properly presenting the resulting degeneracies in parameter fits \citep[see][]{allen07b}, but they require assumptions about parametric functional forms. If completeness corrections are taken into account, it is important to ensure that different surveys use very similar or identical assumptions. In addition, because of the statistical nature of incompleteness corrections, only robust estimates of $CF$ can be obtained since it cannot be determined around which primaries are the companions missed in the survey.

\section{MULTIPLICITY ON THE MS}
\label{sec:ms}

\indent \indent In this section, we review the multiplicity of Population\,I MS stars grouped by mass bins: solar-type stars ($M_\star \approx 0.7$--1.3\,$\,M_\odot$; spectral types F through mid-K, Section\,\ref{subsec:ms_G}), low-mass stars ($M_\star \approx 0.1$--0.5\,$M_\odot$; M0--M6, Section\,\ref{subsec:ms_M}), VLM stars ($M_\star \lesssim 0.1\,M_\odot$; M8 and later, Section\,\ref{subsec:ms_vlm}), intermediate-mass stars ($M_\star \approx 1.5$--5\,$M_\odot$; B5--F2, Section\,\ref{subsec:ms_A}) and high-mass stars ($M_\star \gtrsim 8\,M_\odot$; B2 and earlier, Section\,\ref{subsec:ms_O}). The mass dependence of stellar multiplicity is addressed in Section\,\ref{subsec:trend_mass}.

\subsection{Solar-type stars}
\label{subsec:ms_G}

\subsubsection{Historical overview}

Solar-type stars represent an ideal population to probe stellar multiplicity on the MS. They are frequent enough to yield large, statistically robust samples and, while they are relatively bright and easy to observe, their luminosity contrast with any stellar companion is relatively modest. Solar-type stars were well represented in the samples of bright nearby stars studied over the last century \citep[e.g.,][]{heintz69}, but completeness and sample homogeneity were poor. Even the large scale, uniform and dedicated solar-type multiplicity study conducted by \cite{abt76} had a limited completeness and suffered from the biases inherent to its magnitude-limited nature.

The first modern, volume-limited multiplicity survey of solar-type stars was conducted by \cite{dm91}, who studied 164 objects out to 22\,pc by collating results from many separate studies and obtaining their own new radial velocity measurements. For the last two decades, this study has represented the ``gold standard'' of multiplicity surveys among field stars, against which most subsequent multiplicity surveys compared their results. Nonetheless, significant incompleteness potentially affected the conclusions of \citeauthor{dm91}, prompting \cite{raghavan10} to revisit this topic with a volume-limited sample of 454 star out to 25\,pc. In addition to a much cleaner sample, that survey built on much improved observational methods to reach a very high completeness rate ($\approx 97$\%). This study now represents the most complete multiplicity study of a well-defined sample and likely will remain a standard reference for years to come.

In addition to these global surveys, which consider companions at all orbital periods, valuable results have been obtained in large single-method surveys which enable more uniform detections rates. In the context of field solar-type stars, dedicated surveys for SBs, VBs and common proper motion companions have been conducted \citep[e.g.,][]{mason98, halbwachs03, tokovinin11}.

\subsubsection{Multiplicity frequency}

While \cite{abt76} concluded that ``single stars are infrequent'', both selection biases and survey incompleteness led these authors to significantly overestimate the companion frequency of solar-type stars. Instead of an average of 1.4 companion per primary star, \cite{dm91} and \cite{raghavan10} demonstrated that $CF_{0.7-1.3\,M_\odot}^{\rm MS} = 62 \pm 3\%$. In addition, \citeauthor{raghavan10} estimated that $MF_{0.7-1.3\,M_\odot}^{\rm MS} = 44 \pm 2\%$, i.e., a slight majority of all field solar-type stars are actually single, as is the Sun. Furthermore, \citeauthor{raghavan10} took advantage of their large sample size to identify a dependency of multiplicity on stellar mass. Specifically, they found that that super-solar dwarfs have a marginally higher multiplicity rate than sub-solar dwarfs: $MF_{1-1.3\,M_\odot}^{\rm MS} = 50 \pm 4\%$ vs $MF_{0.7-1\,M_\odot}^{\rm MS} = 41 \pm 3\%$. Similarly, their sample leads to $CF_{1-1.3\,M_\odot}^{\rm MS} = 75 \pm 5\%$ and $CF_{0.7-1\,M_\odot}^{\rm MS} = 56 \pm 4\%$.

\cite{mason98} and \cite{metchev09} have suggested that young ($\sim0.1$--1\,Gyr) solar-type dwarfs have an excess of visual companions over their older counterparts. However, these surveys were skewed towards super-solar targets, reducing the amplitude of the proposed excess. Only a bias-free survey could conclusively demonstrate that young field solar-type stars host more stellar (and substellar) companions than their older MS counterparts. 

\subsubsection{Orbital period distribution}

The distribution of orbital periods for solar-type dwarfs has long been known to be extremely broad, spanning up to 10 decades in orbital periods \citep{luyten30}. While \"Opik's law is a reasonable first approximation, the distribution cuts off at both extremes \citep{heintz69}, although the exact location of the ``breaks'' are somewhat ill-defined \citep[e.g.,][]{poveda07, tokovinin12}. Indeed, the high completeness and large sample size of modern volume-limited samples now allows to confidently exclude \"Opik's law as a good representation of the orbital period distribution over the entire range of periods. On the other hand, the log-normal description that was first proposed by \cite{kuiper35b} has now become a clearly better description of the observed distribution \citep{raghavan10}. The maximum of the orbital period distribution occurs at $\overline{P} \approx 250$\,yr ($\overline{a} \approx 45$\,AU), with a dispersion of $\sigma_{\log P} \approx 2.3$. 

\subsubsection{Mass ratio distribution}

The difficulties associated with the detection of low-mass companions has lead to widely discrepant conclusions regarding the overall mass ratio distribution of solar-type binaries. In turn, it has been proposed that this distribution is flat, bimodal, or monotonically increasing toward low-$q$ systems \citep[][and references therein]{trimble90}. Modern volume-limited surveys have clarified this ambiguous picture. Although \cite{dm91} found a single peak around $q \approx 0.3$, the much higher completeness of the survey by \cite{raghavan10} firmly established that this distribution is instead flat down to $q \approx 0.1$ with the exception of a marginally significant peak at $q \gtrsim 0.95$. We fit a power law to the overall sample from \citeauthor{raghavan10} (adding all three panels in their Figure\,16) and obtain $\gamma_{0.7-1.3\,M_\odot}^{\rm MS} = 0.28 \pm 0.05$. 

At a more detailed level, it is apparent that the mass ratio distribution differs significantly between short- and long-period binaries. Splitting the sample of \cite{raghavan10} at the median orbital period, short-period binaries are characterized by a strong peak at $q \approx 1$ and a slowly declining $f(q)$ function towards low mass ratios, while long-period binaries have a single peak around $q \approx 0.3$, similar to the total distribution found by \cite{dm91}. Specifically, we find that $\gamma_{0.7-1.3\,M_\odot}^{{\rm MS}, \log P \leq 5.5} = 1.16 \pm 0.16$ and $\gamma_{0.7-1.3\,M_\odot}^{{\rm MS}, \log P > 5.5} = -0.01 \pm 0.03$ \cite[see also][]{tokovinin11}, although neither power law is a perfect model of the observed distribution.

One remarkable feature of the mass ratio distribution for solar-type binaries is the dearth of substellar companions, the so-called ``BD desert'', first identified among SBs \citep{marcy00, grether06} and further proposed for VBs \citep[e.g.,][]{leconte10}. We discuss this in the general context of ``extreme'' mass ratio systems in Section\,\ref{subsec:extreme}.

\subsubsection{High-order multiple systems}

Triple and higher-order systems represent $\approx 25\%$ of all solar-type multiple systems, with a distribution of systems with $n$ components that roughly follows a geometric distribution $N(n) \propto 2.5^{-n}$ up to $n = 6$ \citep[see also][]{eggleton08}. However, this distribution is noticeably steeper for non-single systems: $N(n) \propto 3.7^{-n}$.

Not surprisingly, multiple systems are hierarchical. In all systems, the ratio of the orbital periods $P_{\rm long} / P_{\rm short}$ has a minimum value of $\gtrsim 5$ \citep[e.g.,][]{tokovinin97}, consistent with long term stability arguments \citep{eggleton95}. The mass ratio distribution for the short period subsystems is flat up to $q\approx 0.9$, with a strong peak around $q \approx 1$ which is largely responsible for the marginal excess reported in the overall sample of \cite{raghavan10}. On the other hand, the long period subsystems preferentially have $q \leq 0.5$ \citep{tokovinin08, raghavan10}.

More than half of all pairs with $\log P \leq 2$ are members of high-order multiple systems and this proportion rises even further for the shortest-period systems \citep{tokovinin06, allen12}. In addition, the frequency of hierarchical triple systems among wide systems (separation $\geq$1000\,AU) is also high \citep{makarov08}. This suggests that the presence of an outer companion plays an important role in the formation of tight pairs, presumably through energy and angular momentum exchanges. It is plausible that this phenomenon is also related to the formation of equal-mass subsystems. 

\subsection{Low-mass stars}
\label{subsec:ms_M}

\subsubsection{Historical overview}

The high incidence of low-mass stars enables large samples of nearby objects to be searched for stellar companions. After the first attempts at detecting companions via the astrometric perturbation methods \citep[e.g.,][]{lippincott78}, a series of instrumental breakthroughs over the last two decades have helped establish the multiplicity of low-mass stars through radial velocity \citep{marcy89}, direct imaging \citep{skrutskie89} and speckle interferometry \citep{henry90} surveys. Although none of these surveys was volume-limited and their samples only partially overlaped, they allowed \cite{fischer92} to perform a complete analysis of M dwarf multiplicity for 166 stars that has long been used as the low-mass stars counterpart to the \cite{dm91} survey for solar-type stars. More recently, similar-sized volume-limited surveys of low-mass stars have been conducted \citep{reid97b, delfosse04, dieterich12}. In particular, the RECONS consortium\footnote{http://www.recons.org/} has gathered a quasi-complete sample of stars and companions out to $\sim 10$\,pc \citep{henry06}. 

\subsubsection{Multiplicity frequency}

While most studies agree that low-mass stars have fewer companions than solar-type stars, published multiplicity frequencies range from $26 \pm 9\%$ \citep{leinert97} to $42 \pm 9 \%$ \citep{fischer92}. This spread can in part be explained by the small number of companions detected in these surveys, further compounded by significant incompleteness corrections in certain domains. Modern volume-limited surveys, with their near-perfect completeness, draw a much more coherent picture, where $MF_{0.1-0.5\,M_\odot}^{\rm MS}=26\pm3$\% and $CF_{0.1-0.5\,M_\odot}^{\rm MS}=33\pm5$\% \citep{reid97b, delfosse04, dieterich12}. The widely-used higher frequency reported by \cite{fischer92} can be traced back to their analysis of common proper motion binaries, which focused on the earliest spectral types (M3.5 and earlier). Indeed, recent studies have demonstrated that the frequency of wide companions (50--10$^5$\,AU) is a steep function of primary mass \citep{dhital10} so that an average over all spectral types leads to a much reduced companion frequency at large separations. Despite this trend, there is no significant variation of the total multiplicity frequency between early-M and late-M dwarfs since most of their companions are found at separations shorter than 50\,AU. On the other hand, the frequency of spectroscopic companions estimated by \citeauthor{fischer92} has been upheld by more recent surveys \citep{clark12}

\subsubsection{Orbital period distribution}

\cite{fischer92} suggested that a log-normal distribution similar to that found for solar-type binaries could also apply to low-mass stars. As pointed out above, however, the frequency of wide companions was overestimated by \citeauthor{fischer92} and the orbital period distribution must thus be revisited. A log-normal function is still acceptable for separation below 500\,AU, with $\overline{a} \approx 5.3$\,AU and $\sigma_{\log P} \approx 1.3$ based on the latest update to the RECONS sample. At larger separation, an \"Opik-like distribution seems to hold out to 0.2--0.3\,pc for early-M primaries \citep{dhital10}. Considering the small number of companions included in volume-limited samples, and the difficulty to draw a bias-free picture from the other available surveys, it remains unclear whether and how these two distributions are physically connected.

\subsubsection{Mass ratio distribution}

The flat mass ratio distribution found by \cite{fischer92} suffered from potential biases as only primaries in the M0--M3 range and companions significantly above the substellar limit were considered. Nonetheless, including lower mass M-type primaries and all secondaries yields a mass ratio distribution that is flat or slightly declining towards low-$q$ systems. For instance, we have fit a power law distribution to the total sample of \citet[their Figure\,4]{delfosse04} and found that $\gamma_{0.1-0.5\,M_\odot}^{\rm MS} = 0.39\pm 0.23$. However, the situation is more subtle, since the mass ratio distribution is clearly dependent on the primary mass and the binary separation. Like solar-type binaries, short period M-dwarf binaries are significantly biased towards high-$q$ systems \citep{reid97b}. In addition, low-$q$ systems are limited to the earliest M-type dwarfs \citep{dieterich12}. Fitting the RECONS sample, we find that $\gamma_{0.1-0.5\,M_\odot}^{{\rm MS}, a \leq 5\,AU} = 2.7 \pm 1.6$ and $\gamma_{0.1-0.5\,M_\odot}^{{\rm MS}, a > 5\,AU} = -0.3 \pm 0.3$. Similarly, we find that $\gamma_{0.3-0.5\,M_\odot}^{\rm MS} = -0.2 \pm 0.3$ and $\gamma_{0.1-0.3\,M_\odot}^{\rm MS} = 1.9 \pm 1.7$. Finally, we note that BD companions are easier to find around M dwarfs than around solar-type stars. Although dedicated searched have discovered a number of such companions \citep[e.g.,][]{nakajima95, forveille04, radigan08, burningham09}, the frequency of such companions is much lower than that of stellar companions based on volume-limited survey, with only 3 of the 23 companions in the RECONS sample being substellar. 

\subsubsection{High-order multiple systems}

The ratio of multiple to binary systems among low-mass stars is $\approx 21$\% \citep{reid05}, similar to the ratio observed for solar-type stars. Furthermore, the distribution of systems with $n$ components among M dwarfs follows an $N(n) \propto 3.9^{-n}$ for $n \geq 2$, closely matching the distribution for solar-type systems\footnote{See for instance the VLM binary database (http://www.vlmbinaries.org/)}. Finally, high-order multiple systems appear to have a much higher occurrence rate among systems which contain the widest pairs among M-dwarf systems \citep{law10}, suggesting that higher system masses are needed to produce the widest systems. Overall, besides the lower overall multiplicity of M dwarfs compared to solar-type stars, the basic statistics of high-order multiple systems do not seem to differ significantly between these two populations. 

\subsection{VLM stars and substellar objects}
\label{subsec:ms_vlm}

\subsubsection{Historical overview}

VLM objects, even the nearest ones, are typically very faint and therefore difficult targets for both high-resolution imaging and spectroscopic observations. Broad molecular bands and fast rotational velocities further hamper searches for SBs. In his review of BDs, \cite{basri00} only briefly addressed the issue of multiplicity as the first BD binary systems had just been discovered. In the following decade, a slew of high-resolution surveys identified many systems, and thorough reviews of multiplicity in the substellar regime were presented by \cite{burgasser07} and \cite{luhman12}. 

\subsubsection{Multiplicity frequency}

In the absence of a large volume-limited survey for both VBs and SBs, a piece-wise approach is needed to estimate the total multiplicity frequency of VLM stars. The broad range of observed frequency of spectroscopic companions \citep{reid02, guenther03, basri06, blake10, tanner12} is primarily a result of different completeness limits for each survey. Combining all these surveys, 77 objects of spectral type M8 or later have been observed at least three times with high-resolution spectroscopy, revealing 4 confirmed SBs and at least two more candidate binaries. The frequency of SBs ($a \lesssim 1$\,AU) is therefore at least $5.2^{+3.8}_{-1.5}\%$. At larger separations ($a \gtrsim 2$\,AU), high-resolution imaging surveys indicate an observed binary frequency of about 15$\pm$3\% after correction for their inherent Malmquist bias \citep{bouy03, burgasser07}. It is likely that a handful of companions remain undetected in the 1--2.5\,AU range \citep[e.g.,][]{bernat10}. Finally, no more than 2.3\% of VLM objects have a companion in the 40-1000\,AU range \citep{allen07a}.  Combining these various studies yields a total multiplicity fraction for field VLM objects of 20--25\%, consistent with the frequency of $CF_{M_\star \lesssim 0.1\,M_\odot}^{\rm field} =22^{+6}_{-4}\%$ estimated by \cite{allen07b} from a Bayesian analysis of several input surveys. We adopt this latter estimate, which takes into account survey incompleteness. Furthermore, while a few triple VLM systems have been confirmed or suspected \citep[e.g.,][]{artigau11, burgasser12}, too few of these have been discovered to draw a meaningful conclusion. Instead, we simply estimate that $CF_{M_\star \lesssim 0.1\,M_\odot}^{\rm field} \approx MF_{M_\star \lesssim 0.1\,M_\odot}^{\rm field}$. While the studies discussed so far generally focus on late-M or L-type objects, recent surveys found a similar multiplicity frequency for the coldest (T-type) BDs. Even the higher frequency of visual companions among ``transition'' objects (late-L to early-T dwarfs) can be explained by an enhanced Malmquist bias \citep{burgasser10}. 

\subsubsection{Orbital period distribution}

While too few VLM binaries are known to derive robust estimates, their orbital period distribution is much narrower than that of higher mass objects. Most companions are found with projected separations in the 1--10\,AU range \citep[see][]{burgasser07}. Only a handful of systems with projected separations larger than 50\,AU are known \citep{billeres05, burningham10, dhital11} and this near-absence of wide systems cannot be accounted for by incompleteness \citep{allen07b}. \cite{maxted05} and \cite{allen07b} have used statistical analyses to derive (truncated) log-normal analytical descriptions of the overall orbital period distribution. Both studies agree regarding the peak of the separation distribution ($\overline{a} \approx 4$--7\,AU), but differ on the width of the distribution. It is probable that an intermediate width ($\sigma_{\log P} \approx 0.4$--0.5) is a good match to all observations. However, the orbital period distribution may follow a different analytical prescription and may even be asymmetric. More thorough searches for SBs are needed to settle this issue.

\subsubsection{Mass ratio distribution}

The mass ratio distribution for VLM stars\footnote{For substellar objects, estimated masses depend on the assumed age of the objects and are therefore not as precise as for hydrogen-burning stars. However, the uncertainty applies similarly to both components of a presumed coeval binary, so that absolute uncertainties in $q$ are substantially smaller than on the masses of individual components.} is heavily skewed toward equal mass systems \citep[see][]{burgasser07}. The vast majority of known systems have $q \geq 0.7$ and the most ``extreme'' systems have $q \approx 0.5$ \citep[e.g.,][]{biller06, kasper07}. Several studies have attempted to reproduce the observed mass ratio distribution by assuming an intrinsic power law distribution and taking into account selection and incompleteness effects \citep{burgasser06, allen07b, liu10}. While they all agree that $\gamma_{M_\star \lesssim 0.1\,M_\odot} \lesssim 0$ can be firmly excluded, the estimated range of estimated values is very broad. This diversity is most likely the result of methodological differences. The shallow index found by \cite{allen07b} is unlikely to be appropriate as it leads to the prediction of many binaries with $q \leq 0.7$, which is at odds with deep imaging surveys. Instead, \cite{burgasser06} argues that the sharp decline in the observed distribution of mass ratios as $q$ decreases is only modestly amplified by the Malmquist bias. While this issue is not entirely settled, we adopt the steep distribution proposed by \citeauthor{burgasser06}, $\gamma_{M_\star \lesssim 0.1\,M_\odot} = 4.2 \pm 1.0$.

\subsection{Intermediate-mass stars}
\label{subsec:ms_A}

\subsubsection{Historical overview}

Intermediate-mass stars are much more challenging for multiplicity studies than lower-mass stars. The steep mass-luminosity relationship makes low-mass companions hard to detect, and given typical distances, there is an un-probed gap between SBs and VBs. Further complicating the picture is the presence, among nearby field A dwarfs of some $\sim 30\%$ of chemically peculiar stars. Some of these categories owe their peculiarity to multiplicity \citep{abt65}, inducing severe selection biases. In this review, we consider A stars as a unique population, averaging over the various sub-samples.

Searches for SBs among intermediate-mass stars have been ongoing for several decades \citep[e.g.,][]{abt73, carrier02, carquillat07}. Systematic searches for visual companions have been largely incomplete for a long time, although some of the studies listed above have attempted to include wider companions. The efforts to identify low-mass visual companions have accelerated in recent years as the presence of BD, and more recently planetary-mass, companions is investigated \citep[e.g.,][]{ivanov06, balega11}. 

In parallel to these studies of field stars, another avenue to probe the multiplicity of intermediate-mass stars is to consider rich stellar associations in the process of dissipating in the field. By an age of 5\,Myr, intermediate-mass stars already have reached the zero-age MS. The Scorpius-Centaurus OB association is the best studied region to date, both for SBs \citep[e.g.,][]{brown97} and visual companions \citep{shatsky02, kouwenhoven07a}, which allowed \cite{kouwenhoven07b} to perform a systematic analysis to derive the global multiplicity properties of intermediate-mass stars in that population.

\subsubsection{Multiplicity frequency}

The observed frequency of SBs among field intermediate-mass stars lies in the 30--45\% range \citep{abt83}. In the Sco-Cen OB association, the fraction of SBs is at least 30\%. The most thorough search for visual companions to A stars is the ongoing VAST survey \citep{derosa11}. In the separation range 50--2000\,AU, over which the survey is complete down to the substellar limit, they find a frequency of companions of 40$\pm$4\%, although many of these companions have $q \leq 0.1$ and some of them may be unrelated background stars (R. De Rosa, priv. comm.). In the Sco-Cen OB association, \cite{kouwenhoven05} found a frequency of visual companions of 37\%. Considering the limitations of current surveys, the Sco-Cen population provides the most robust estimate of the multiplicity of intermediate-mass stars. Current observations support the hypothesis that field stars have similar properties. Based on the various surveys listed above, we conclude that $MF_{1.5-5\,M_\odot}^{\rm MS} \geq 50$\%, although incompleteness precludes robust estimates of the frequency of high-order multiplicity. In addition, the statistical analysis of \cite{kouwenhoven07b} resulted in $CF_{1.5-5\,M_\odot}^{\rm MS} = 100 \pm 10$\%.

\subsubsection{Orbital period distribution}

The orbital period distribution for field intermediate-mass multiple systems extends from less than 1\,d to projected separations of well beyond $10^4$\,AU \citep{abt83}. Contrary to lower-mass field binaries, however, the orbital period distribution for A stars appears to be bimodal, with a peak among SBs around $P \sim 10$\,d \citep[e.g.,][]{carquillat07} and one for VBs around 350\,AU (R. De Rosa, priv. comm.). In addition, there is marginal evidence for a dip around $P \sim 300$\,d, with a tertiary peak or a broad plateau for separations in the 1--50\,AU \citep{budaj99}. Even though the observed distribution seems more complicated, \cite{kouwenhoven07b} tried both a power law and a log-normal representation for the period distribution of the Sco-Cen intermediate-mass binaries with moderate success. In the former case, \"Opik's law is favored, while in the latter formalism, they find $\sigma_{\log P} \gtrsim 2.3$. However, neither is a perfect description of the observed distribution and no simple analytical representation matches all observations.

\subsubsection{Mass ratio distribution}

The mass ratio distribution of field intermediate-mass stars remains elusive as a result of incompleteness and selection biases, which are largely responsible for the early claims of a strong preference for high-$q$ systems \citep[e.g.,][]{abt85}. Taking those limitations into account, it now appears that the mass ratio distribution for SBs is either relatively shallow \citep[e.g., $\gamma = -0.3 \pm 0.2$;][]{carquillat07} or can be described by a broad Gaussian distribution centered around $q \approx 0.4$ \citep{carrier02, vuissoz04}. Among visual companions, the VAST survey suggests a mass ratio distribution $\gamma \approx -0.6$ that is roughly consistent with that found for SBs (R. De Rosa, priv. comm.). Finally, we note that a similar mass ratio distribution was proposed by \cite{shatsky02} and \cite{kouwenhoven07b} for the Sco-Cen association. The latter study derived $\gamma_{1.5-5\,M_\odot}^{\rm UpSco} = -0.45 \pm 0.15$. Since this analysis includes both SBs and VBs, we adopt it as representative of all intermediate-mass multiple systems.

\subsection{High-mass stars}
\label{subsec:ms_O}

\subsubsection{Historical overview}

The multiplicity of high-mass stars has long been the subject of intense scrutiny \citep[][and references therein]{zinnecker07}. However, their large distances from the Sun, large rotational velocity and the extremely high brightness contrast requirements conspire to draw an incomplete picture \citep{sana11}. In addition, most high-mass stars are found in their birth environment, massive clusters or OB associations, and only $\sim20$\% of all nearby O stars are found in the field, many of which are thought to be ejected ``runaway'' stars \citep[e.g.,][]{dewit04, chini12}. As a result, studying only ``field stars'' is likely to yield a very biased picture.

High-mass SBs have been studied in all environments for several decades \citep{garmany80, abt90, chini12}. Individual clusters and OB associations, despite smaller sample sizes, are more appropriate for a uniform characterization of multiple systems \citep{sana09, kiminki12}. Furthermore, it has recently become possible to probe the multiplicity of high-mass stars in the low-metallicity environment of the Large Magellanic Cloud \citep{sana12b}. Surveys for visual companions to high-mass stars have been conducted both for ``all-sky'' samples \citep[e.g.,][]{turner08, mason09} and single clusters/associations \citep[e.g.,][]{duchene01, peter12}. 

\subsubsection{Multiplicity frequency}

The remarkably broad range of estimated frequency of spectroscopic companions among high-mass stars is due to a combination of small sample sizes, selection biases and incompleteness. Most recently, \cite{sana12} estimated a bias- and completeness-corrected frequency of spectroscopic companions of 70$\pm$9\% out to $P \sim 3000$\,d and down to $q \approx 0.1$. Early-B stars have a lower SB frequency than their higher-mass counterparts \citep[$52 \pm 4 \%$ for B2-B6 primaries;][]{chini12}. Due to the diversity of wavelengths and observing techniques, imaging surveys yield an even broader range of frequencies. A consensus has emerged that the frequency of visual companions over two decades in projected separations is $\sim 45 \pm 5$\% \citep[e.g.,][]{turner08, peter12}. Incompleteness in these surveys is modest down to $q \approx 0.1$ overall but very uncertain below this limit. There is suggestive evidence that the frequency of visual companions decreases with the mass of primary \citep{preibisch99, peter12}. Interpolating between the populations of SBs and VBs suggests a ``missed'' companion frequency in the 10--20\% range, in agreement with the small survey by \cite{nelan04}. This leads to $CF_{M_\star \gtrsim 16\,M_\odot}^{q \gtrsim 0.1} \approx 130 \pm 20$\% and $CF_{8-16\,M_\odot}^{q \gtrsim 0.1} \approx 100 \pm 20$\%. The latter is substantially lower than the estimate from \cite{abt90} but the large incompleteness correction of that survey probably explains this difference. In addition, the number of high-mass stars that are truly single stars is remarkably low. Considering only confirmed companions, $MF_{M_\star \gtrsim 16\,M_\odot} \geq 80$\% \citep{mason09, sana11, chini12}. The situation is less well constrained for early-B stars, and $MF_{8-16\,M_\odot} \geq 60$\% \citep{abt90} is a conservative lower limit since many tight visual companions were missed in that survey. High-mass stars in rich clusters and OB associations have essentially identical multiplicity frequencies, which in turn is significantly higher than that of field and runaway stars \citep{turner08, mason09, chini12}. 

\subsubsection{Orbital period distribution}

The relative lack of SBs with published orbital solutions has long impaired definitive analyses of the orbital period distribution for high-mass multiple systems. The high frequency of short-period binaries proposed early on \citep{garmany80, abt90} was potentially subject to a strong selection bias, but has been confirmed by modern surveys, which find a pile-up of binaries with $4\,{\rm d} \leq P \leq 8\,{\rm d}$ \cite[e.g.,][]{sana12}. Power law fits suggest a slowly-declining ($\alpha \approx -0.5$) distribution out to $P = 3000$\,d for both Galactic and Large Magellanic Cloud high-mass binaries \citep{sana12,sana12b}. However, extrapolating even a moderately steep distribution results in far too few visual companions, suggesting a more complex functional form. One possibility to match all observations combines two independent distributions: a population of short period binaries ($\log P \lesssim 1$, 30\% of all high-mass stars) and a power law period distribution extending out to $\gtrsim 10^4$ AU. While an \"Opik-like law for the second population is compatible with available observations \citep{kiminki12b}, imaging surveys which repeatedly  find a ``peak'' at the shortest probed angular separations rather support a slowly declining distribution instead \citep{nelan04, mason09, sana11b, peter12}.

\subsubsection{Mass ratio distribution}

Limited sensitivity to even intermediate-mass companions has long plagued studies of the mass ratio distribution of high-mass stars. Furthermore, a large fraction of their companions are still in the PMS phase, resulting in model-dependent mass estimates. Both \cite{kiminki12b} and \cite{sana12} concluded that the intrinsic distribution of mass ratios for high-mass SBs is essentially flat \citep[e.g., $\gamma_{M_\star \gtrsim 16\,M_\odot}^{P \leq 3000\,d, q \geq 0.1} = -0.1 \pm 0.6$ for][which we adopt here]{sana12}. While both studies find a peak around $q \approx 0.8$, they rule out the presence of a separate population of equal-mass (so-called ``twin'') systems among SMC eclipsing binaries \citep{pinsonneault06}. Among VBs, \cite{peter12} concluded that a flat distribution and a broad Gaussian distribution centered on $q \sim 0.45$ match equally well their observations of high-mass stars in the Carina region. On the other hand, surveys of the Orion Nebula Cluster \citep[ONC;][]{preibisch99} and NGC\,6611 \citep{duchene01} found a strong preference for low-mass companions ($q \leq 0.5$), although some of these systems may not be physically bound. Combining these three samples and fitting a power law to the resulting distribution yields $\gamma_{M_\star \gtrsim 16\,M_\odot}^{a \geq 100\,AU} = -0.55 \pm 0.13$, indeed favoring lower-mass companions but not significantly different from the distribution found for SBs down to $\sim 1\,M_\odot$. Lower mass companions remain mostly out of reach of present surveys but the upcoming generation of high-contrast, high-resolution imaging instruments will soon provide a first foray into this category of ``extreme'' systems ($q << 0.1$). 

\section{MULTIPLICITY OF OLD AND YOUNG STARS}
\label{sec:time}

\indent \indent In this section, we consider multiplicity properties of stellar populations of old MS stars (Population\,II stars) and at increasing earlier stages of stellar evolution, from open clusters to the youngest embedded protostars. The time evolution of multiplicity is addressed in Section\,\ref{subsec:trend_time}.

\subsection{Population\,II stars}
\label{subsec:popII}

\indent \indent The multiplicity properties of Population\,II stars can be used to probe the star formation process in low metallicity environments. While early studies led to conflicting conclusions regarding a possible deficit of companions to Population\,II stars \citep{oort26, jaschek59, abt87}, a more coherent picture has recently emerged thanks to instrumental improvements and larger and better-defined samples of low-metallicity dwarfs and high-velocity halo stars \citep[e.g.,][]{carney94}. 

SBs among Population\,II stars appear to be equally frequent and to share very similar properties with their higher-metallicity counterparts \citep{goldberg02, latham02}. On the other hand, the frequency of VBs ($a \ge 10$\,AU) is markedly lower than for solar-metallicity stars \citep{allen00, zapatero04, lodieu09}, though perhaps not for K-M dwarfs \citep{jao09}. As a result, the total multiplicity frequency of Population\,II stars is $CF_{0.7-1.3\,M_\odot}^{Pop\,II} = 39 \pm 3\,\%$ and $CF_{0.1-0.6\,M_\odot}^{\rm Pop\,II} = 26 \pm 6$\,\% \citep[][respectively]{rastegaev10, jao09}.

The distribution of orbital period among solar-mass Population\,II multiple systems is characterized by a narrow peak in orbital period around $\log P \approx 2$--3, accompanied by a long-period tail out to $\sim 10^4$\,AU that follows roughly \"Opik's law \citep{allen00} and a marked dearth of systems with orbital periods shorter than $\log P \approx 1$ \citep{latham02, rastegaev10}. The secondary peak at $\sim 500$\,AU proposed by \cite{zinnecker04} has not been confirmed by \cite{rastegaev10}, casting doubt on the its existence. Too few low-mass Population\,II binaries are known to robustly estimate their orbital period distribution, though they may represent a broader distribution than for solar-metallicity stars (Jao et al. 2009). Finally, while the mass ratio distribution of solar-type Population\,II binaries is still poorly determined, companion masses down to the substellar limit are found in roughly uniform distribution \citep{zapatero04}. 

\subsection{Open clusters}
\label{subsec:open_clusters}

\indent \indent Members of open clusters represent young (typically  50\,Myr to 1\,Gyr) and uniform populations that will eventually dissipate in the galactic field. These clusters are dynamically old, in the sense that their age is much larger than their crossing time, and wide binaries have long been disrupted \citep[e.g.,][]{kroupa95, parker09}. Despite the advantage of large and homogeneous samples, multiplicity studies of open clusters are surprisingly scattered in the literature, with little overlap between samples. The best studied open clusters, in which both SBs and close VBs have been searched for, are the nearest ones ($D \leq 200$\,pc): $\alpha$\,Per, the Pleiades, Praesepe and the Hyades. No dramatic difference has been identified between these clusters in terms of their multiplicity properties so that a composite average can be constructed to draw a global picture. 

\subsubsection{Multiplicity frequency}

We first address the multiplicity of solar-type stars, which are the best-studied in open clusters. Most survey for SBs in open clusters agree on a 13--20\% frequency of spectroscopic companions \citep[e.g.,][]{prosser92, mermilliod08b, griffin12}. Although adaptive optics surveys rely on substantial incompleteness correction for the closest companions ($\rho \lesssim 1''$), they are more sensitive to low-mass companions than speckle surveys \citep{mason93a, mason93b, patience98, patience02} and their companion frequency of $\sim 25$\% over 1.5--2 decades in separation is more robust.

Combining SB and VB surveys therefore yields an observed companion frequency of open clusters of $\approx 40\%$ for solar-type stars, although some companions most likely still escape detection. Taking advantage of the range of distances to various clusters, \cite{patience02} found a total companion frequency of $48 \pm 5$\% for all separations up to 580\,AU and with mass ratios $q \gtrsim 0.25$. Assuming a mass ratio distribution similar to that of field FGK dwarfs ($\gamma \sim 0$), the total frequency is therefore $CF_{0.7-1.3\,M_\odot}^{\rm open\, cluster} \sim 65$\%. Independently, photometric studies have suggested multiplicity frequencies in the 65--70\% range for solar-type stars \citep{kahler99, converse08}. 

Low-mass cluster members have only been systematically studied via spectroscopic and imaging surveys in the nearby Hyades cluster \citep{reid97, reid00}. Combining these two surveys leads to $CF_{0.1-0.5\,M_\odot}^{\rm open\,cluster} \approx 35 \pm 5$\%. Finally, VLM stars and BDs in open clusters have only been searched for visual companions \citep{martin03, bouy06}. Extrapolating the observed frequency of visual companions assuming an overall Gaussian orbital period distribution, \cite{gizis95} suggested that $CF_{M_\star \lesssim 0.1\,M_\odot}^{\rm open\,cluster} = 27\pm16\%$. Photometric studies, which may be affected by a mass segregation selection effect, broadly support these results despite some significant scatter \citep{pinfield03, lodieu07, converse08, boudreault12}.

\subsubsection{Orbital period distribution}

Although no single cluster has been studied over the entire range of orbital periods, it is possible to estimate the average orbital period distribution in open clusters in a piece-wise manner using the various surveys discussed above. The distribution for solar-type stars is broad and unimodal over the range $0 \leq \log P \leq 7$ \citep{patience02}. The second peak at $\log P \leq 1$ proposed by \cite{griffin12} likely stems from the fact that their sample includes intermediate-mass stars.

The median separation of $\sim 4$\,AU estimated by \cite{patience02} based on a log-normal fit, somewhat tighter than among field stars, may be a consequence of incompleteness among VBs. Indeed, the frequency of VBs included in their analysis by \citeauthor{patience02} appears to be underestimated \citep{bouvier97, bouvier01, morzinski12}, suggesting that the median separation is significantly larger. It is therefore probable that the orbital period distribution of solar-type binaries in open clusters is indistinguishable from that of field stars. Unfortunately, chance projection of unrelated members and background stars prevent any analysis of systems wider than $a \sim 1000$\,AU in open clusters.

\subsubsection{Mass ratio distribution}

The distribution of mass ratios among solar-type binaries in open clusters is essentially flat for both SBs \citep[e.g.,][]{goldberg94, mermilliod99, bender08} and VBs \citep{patience02}. In addition, we note that substellar companions to solar-type stars in open clusters are rare \citep{metchev09, itoh11}, although a few have been recently discovered \citep{geissler12, rodriguez12}. On the other hand, multiple systems are increasingly skewed toward high-$q$ for low-mass \citep{reid97, reid00} and VLM \citep{martin03, bouy06, lodieu07} members of open clusters.

\subsection{Young nearby associations}

\indent \indent The past decade has seen the discovery of several ``moving groups'', i.e., groups of co-moving, coeval young stars that are unbound and not associated with a recognizable site of star formation \citep[see][]{torres08}. The ages of these groups ($\sim 77$--100 Myr) make them a valuable bridge between the sites of ongoing star formation on one hand and open clusters and the field population on the other. Most of these groups are small ($N \lesssim 50$), however, so they individually do not provide robust statistical constraints on the full binary population. Furthermore, many of the known binaries were discovered serendipitously rather than in statistically robust surveys. The few analyses to date suggest that they are broadly consistent with what is seen in the field and in young star-forming regions \citep[e.g.,][]{brandeker03, evans12}, except for an anomalously low frequency of wide binaries ($a \geq 20$\,AU) in the $\eta$\,Cha moving group \citep{brandeker06}. Because of their youth and close proximity to the Sun ($d \leq 50$ pc), members of these associations have been subject to surveys aiming primarily at planet discovery \citep[][and references therein]{chauvin10}. These have revealed a small number of planetary-mass companions \citep{chauvin04, lagrange09} and numerous BDs \citep[e.g.,][]{lowrance99, neuhauser04, wahhaj11, bowler12}, as well as providing robust constraints on the existence and extent of the purported BD desert \citep{evans12}. A uniform analysis of all these results is still missing.

\subsection{PMS stars}
\label{subsec:TTS}

\indent \indent We now turn to multiplicity surveys targeting populations of PMS stars with ages in the 1--5\,Myr range and with stellar mass $M_\star \lesssim 2\,M_\odot$, collectively called T\,Tauri stars (TTS), or Class\,II/III sources depending on whether they host a protoplanetary disk \citep{lada87}. Multiplicity surveys have targeted TTS located in a broad range of environments: dense stellar clusters (such as the ONC and IC\,348), young OB associations (e.g., Sco-Cen), and ``loose'' (so-called T) associations (like the Taurus-Auriga or Chamaeleon star-forming regions). The latter populations typically contain many less stars ($\sim 100$) than clusters and OB associations and only a handful of intermediate-mass stars. The closest of these populations are located $\sim 150$\,pc away from the Sun. 

\subsubsection{Historical Overview}

Even though the existence of close pairs of TTS was noted very early with seeing-limited observations \citep[][]{joy44, herbig62}, the multiplicity of TTS was only addressed by dedicated surveys in the 1990s. Since then, thorough multi-wavelength surveys of nearby star-forming region have allowed the definition of large, complete samples for multiplicity studies \citep[e.g.,][]{hbc88, luhman12}. In addition to seeing-limited surveys \citep[][]{reipurth93} and mining of all-sky databases \cite[][]{kraus09}, closer companions have been searched with a variety of high-angular resolution techniques, down to 2--3\,AU resolution. In more or less chronological order, speckle interferometry \cite[][]{ghez93, leinert93}, lunar occultation \citep{simon95}, Hubble Space Telescope imaging \cite[][]{padgett97}, adaptive optics \cite[][]{duchene99} and aperture masking \cite[][]{kraus08} have all been used in this context. The search for tighter companions remains challenging as strong and variable emission lines hamper radial velocity analyses. Nonetheless, a small number of SBs are known \citep[e.g.,][]{mathieu89}. The only systematic surveys published to date \citep{melo03, nguyen12} found a companion frequency similar to that of field stars but yielded no new orbital solutions. In the remainder of this section, we only discuss VB surveys, which have been much more extensively studied.

\subsubsection{Multiplicity Frequency}

Binary surveys of PMS stars in T\,associations generally have measured companion frequencies that are roughly twice as high as among solar-type MS stars \citep[e.g.,][]{duchene99b}. This led to the widespread meme that all young stars are born as multiples. The largest complete surveys to date have been conducted in Taurus and Upper Scorpius \citep{kraus08, kraus11}. In Taurus, for instance, $CF_{0.7-2.5\,M_\odot}^{\rm Class\,II/III, Tau} = 64^{+11}_{-9}$\% and $CF_{0.25-0.7\,M_\odot}^{\rm Class\,II/III, Tau} = 79^{+12}_{-11}$\% over the separation range 3--5000\,AU. Despite these high frequencies, it appears that $\sim 1$/4--1/3 of all systems consist of a single star, revealing the prevalence of high-order multiples. \cite{kraus11} noted that the distribution of degrees of multiplicity was broadly consistent with a purely Poissonian distribution, $N(n) \propto e^{-n}$, and hence that adding additional companions resembles a stochastic process. This distribution resembles the geometric distribution seen for the field (Section 3.1.5; Eggleton \& Tokovinin 2008).

At the other extreme, numerous studies have shown that the multiplicity frequency in young dense clusters is much lower. This applies to the locally extreme ONC \citep{petr98, kohler06}, but also to other young clusters in the Orion complex \citep{beck03} as well as to IC\,348 in the Perseus cloud \citep{luhman05}. In the ONC, where the largest sample of targets has been assembled, \cite{reipurth07} found $CF^{\rm Class\,II/III, ONC} = 8.8 \pm 1.1$\% over a decade in projected separation, a factor of $\sim 2.5$ lower than found in Taurus over the same range. Although it has long been proposed that multiplicity is a smooth function of the stellar density, the statistical precision of current surveys in intermediate regions, such as the Ophiuchus cloud, do not conclusively support his scenario \citep{marks12, king12}.

Although they long remained unaccessible, several surveys have targeted BDs and VLM stars in recent years \citep[][and references therein]{biller11}. Collectively, they demonstrated that most young low-mass objects have no companions at projected separations $\geq$5--10 AU. Furthermore, \cite{kraus12} performed a combined analysis of all published surveys and showed that the binary frequency declines precipitously from 0.5 $M_{\odot}$ to below the substellar boundary, with $CF_{M_\star \lesssim 0.1\,M_{\odot}}^{\rm Class\,II/III} < 11\%$ (50\% confidence)

\subsubsection{Orbital Period Distribution}

As for the field population, both a log-normal and power law have been used with reasonable success to describe the distribution of orbital periods among TTS for the last two decades. The inability to distinguish between these distributions results in part from the limited range probed by observations. Recent studies suggest a separation distribution that follows \"Opik's law out of $\sim 5000$\,AU for solar-type stars and slowly declining for lower-mass stars \citep[e.g.,][]{kraus11}. As a result, the global distribution is consistent with both a very broad log-normal, with $\sigma_{\log P} \gtrsim 2$ \citep{kraus11}, or a slowly declining power law \citep[e.g.,][]{king12b}. The latter conclusion also matches the results of \cite{kouwenhoven07b} for intermediate-mass multiple systems in Upper Scorpius. One important caveat is that the binary population in dense clusters drops sharply at projected separations of $\gtrsim$200\,AU \citep{scally99, reipurth07}. 

The orbital period distribution of binary TTS appears to be truncated, with very few systems outside of a maximum separation that varies smoothly with stellar mass \citep[][and references therein]{kraus12}. The truncation radius appears to be $\sim 300$\,AU for stars in the 0.25--0.5\,$M_\odot$ and as small as $\sim25$\,AU for the lowest mass TTS. However, the paucity is not complete: a small number of ``unusually wide'' pairs have been identified \citep[][and references therein]{konopacky07, kraus11}. These are discussed in a broader context in Section\,\ref{subsec:wide}.

Finally, we note that the distribution of orbital separations is skewed towards tighter systems in disk-less Class\,III objects relative to Class\,II objects which host circumstellar disks. This is likely a result of disk evolution, since tidal forces in tight binaries ($a \lesssim 50$\,AU) can severely truncate disks and result in much shorter disk lifetime \citep{cieza09, kraus12b}.

\subsubsection{Mass Ratio Distribution}

Given the shallow mass-luminosity relation for PMS stars, imaging surveys are quite sensitive to low-mass companions, routinely reaching $q \sim 0.1$ and more recently probing down to $q \le 0.01$. However, uncertainties in PMS evolutionary models \citep{hillenbrand04} and the possibility of circumstellar dust emission and differential extinction could work to smooth out and/or bias the genuine mass ratio distributions, particularly for photometrically estimated mass ratios \citep{woitas01}. Nonetheless, in the absence of complete spectroscopic follow-up campaigns \citep[e.g.,][]{white01, hartigan03}, this is the only available tool in most cases. 

Most surveys of solar-type binary TTS have found mass ratio distributions that are roughly flat \citep{lafreniere08, kraus08, kraus11}. Using a power law parametrization for the mass ratio distribution, \cite{kraus11} found $\gamma_{0.25-2.5\,M_\odot}^{\rm Class\,II/III} = 0.2 \pm 0.2$ for the entire Taurus population, albeit with a marginally significant trend with stellar mass, whereby lower mass systems preferentially have $q \gtrsim 0.5$. This trend is further expanded by considering the lowest mass binaries: \cite{kraus12} found $\gamma_{0.07-0.12\,M_\odot}^{\rm Class\,II/III} \sim 1$. Due to the paucity of truly substellar binary systems, it is uncertain whether this trend continues through the substellar regime. We note, however, that there is a small fraction of objects that seem to host binary companions that are much less massive than the primary, and may also have unusually wide orbital radii \citep[e.g.,][]{luhman09}. These are further discussed in Section\,\ref{subsec:wide}. At the other extreme, intermediate-mass binaries have a small but significant preference for low-$q$ systems \citep{kouwenhoven07b}.

\subsection{Embedded protostars}
\label{subsec:protostars}

\indent \indent Over the last decade, increasing attention has been devoted to studying the multiplicity of embedded protostars, objects in the earliest phases of stellar evolution \cite[e.g.,][]{duchene07b}. These represent the best chance of probing the formation mechanism of multiple systems at its origin, before subsequent internal and external dynamical evolution can affect unstable systems. However, protostars also pose serious challenges to companion searches. By definition, they are still embedded in the remnants of their nascent core and hence are best studied at infrared and radio wavelengths, where high-resolution techniques became available only recently. Furthermore, the mass of the central objects are extremely difficult to estimate.
 
Protostars are traditionally categorized based on their spectral energy distribution. Class\,0 sources are the least evolved objects, with more mass in circumstellar material, mostly in the form of a dense envelope, than in the central object \citep{andre93}. In contrast, the envelope mass of Class\,I sources is lower than that of the central object. We discuss both categories separately below.

\subsubsection{Class\,I protostars}

Although they are deeply embedded, Class\,I sources are bright near- and mid-infrared sources. To date, about 350 Class\,I sources have been observed with seeing-limited imaging \citep{haisch04, connelley08} and, for about 100 of them, with high-resolution imaging techniques \citep{duchene07, connelley09}. Until a global statistical analysis of these various surveys is performed, we estimate the multiplicity of Class\,I in the following manner. First of all, we only consider sources within 500\,pc of the Sun and companions with flux ratios $\Delta L \leq 4$\,mag to limit incompleteness issues. The direct imaging survey of \cite{connelley08} yields 27 companions in the 200--2000\,AU range out of 136 targets, while the adaptive optics surveys of \cite{duchene07} and \cite{connelley09} found 15 companions between 50 and 200\,AU to 88 targets (with additional companions at shorter separations for the closest targets). Combining these two fractions, we find a total frequency of companions to Class\,I sources over the 50--2000\,AU separation range of $CF^{\rm Class\,I} = 36.9 \pm 5.2$\%. The associated distribution of projected separations is consistent with \"Opik's law out to 5000\,AU although chance projection of unrelated protostars could be non-negligible at the largest separations \citep{connelley08}. 

There is marginal evidence for an environment dependency among Class\,I protobinaries in that the distribution of separations for Orion shows a marked peak in the 60--200\,AU range not seen in other nearby star-forming regions \citep{duchene07, connelley09}. In this context, it is important to emphasize that Class\,I sources are distributed through the various molecular clouds in the Orion complex (a majority are in the L1641 cloud) and none belongs to the ONC. In other words, even the Orion Class\,I sources studied to date were born in dynamically quiet environments, so that any difference in that population is more likely to be a consequence of some other environmental difference than dynamical evolution. There is also evidence for a temporal evolution of the properties of protobinaries, as the most embedded, and presumably youngest, systems almost exclusively possess wide companions (1000\,AU and wider), the frequency of which rapidly declines as one considers more evolved protostars \citep{connelley08}. 

The search for much tighter companions among Class\,I protostars is remarkably challenging as radial velocity measurements are hampered by extremely weak photospheric features. Based on 2 or 3 measurements per systems \citep{viana12} or comparing a protostar radial velocity to that of gas in its parent cloud \citep{covey06}, small-scaled surveys suggest that some systems tighter than a few AU have already formed by the Class\,I phase, but their interpretation remains ambiguous.

\subsubsection{Class\,0 protostars}

The rarity of Class\,0 sources makes it necessary to group all observed targets in a single sample to achieve meaningful statistics, precluding studies of environmental dependencies. In addition, high-resolution observations at near-infrared and centimeter wavelengths are probing scattered light and free-free emission \citep{reipurth00, reipurth04}, respectively, neither of which can be used to unambiguously detect companions. Instead, the optimal observational probe of Class\,0 multiplicity is therefore millimeter (broadly speaking, 850\,$\mu$m--7\,mm) interferometric mapping, even though the linear resolution typically achieved with such observations is only $\sim 100$\,AU. Furthermore, clustering on scales of 1000\,AU to 0.1\,pc prevents studies of the widest systems \citep[e.g.,][]{peretto07, carrasco12}. 

Early multiplicity studies of Class\,0 protostars suggested that they have a very high multiplicity frequency although their target selection was mostly haphazard and potentially biased \citep[e.g.,][]{looney00, reipurth00}. Recent high-resolution millimeter surveys, which primarily focused on the disk/envelope structure, reported low companion frequencies \citep{maury10}, albeit with extremely small sample size. Adding to the sample surveyed by \citeauthor{maury10}, \cite{enoch11} concluded that only 1 out of 11 targets had a companion within 2000\,AU (albeit with minimum separations probed ranging from 50 to 400\,AU). However, a much larger number of Class\,0 sources have been imaged at high spatial resolution, leading to the discovery of several other companions in that separation range, thereby casting doubts on their apparent rarity \cite[e.g.,][]{looney00}. Further complicating the picture is the possibility that some continuum point sources could be related to outflow features instead of being bona fide companions \citep[e.g.,][]{maury12}, a possibility that can only be clarified with follow-up observations. A global re-analysis of all high-resolution millimeter maps of Class\,0 sources in the context of multiplicity is crucially missing but beyond the scope of this review. For now, we simply note that the multiplicity of Class\,0 sources is not statistically significantly lower than that of the more evolved Class\,I protostars in the $\sim 100$--2000\,AU range. 

\section{OVERALL TRENDS AND ``EXTREME'' SYSTEMS}
\label{sec:trends}

\indent \indent We now gather the results discussed in the previous two sections to highlight the main trends associated with multiplicity as a function of primary mass and age. We then discuss in more detail several remarkable categories of companions: twin companions, extreme mass ratio systems and extremely wide systems. Possible physical implications of empirical trends are discussed in Section\,\ref{sec:implications}.

\subsection{Multiplicity as a function of stellar mass}
\label{subsec:trend_mass}

\indent \indent In this section, we focus on the mass-dependence of multiplicity properties on the MS (see Section\,\ref{sec:ms} and Table\,\ref{tab:multMS}). With continuously improving completeness and larger sample sizes, our empirical knowledge will soon allow decisive tests of the predictions from star formation theories.

\begin{table}
\caption{\label{tab:multMS}Multiplicity properties for Population\,I MS stars and field BDs}
\centering
\begin{tabular}{cccc}
\hline
Mass Range & Mult./Comp. & Mass Ratio & Orbital Period\\
 & Frequency & Distribution & Distribution \\
\toprule
\multirow{2}{*}{$M_\star \lesssim 0.1\,M_\odot$} & $MF = 22^{+6}_{-4}\,\%$ & \multirow{2}{*}{$\gamma = 4.2 \pm 1.0$} & Unimodal (log-normal?) \\
 & $CF = 22^{+6}_{-4}\,\%$ & & $\overline{a} \approx 4.5$\,AU, $\sigma_{\log P} \approx 0.5$ \\
\hline
\multirow{2}{*}{$0.1\,M_\odot \lesssim M_\star \lesssim 0.5\,M_\odot$} & $MF = 26\pm3\,\%$ & \multirow{2}{*}{$\gamma = 0.4 \pm 0.2$} & Unimodal (log-normal?) \\
 & $CF = 33\pm5\,\%$ & & $\overline{a} \approx 5.3$\,AU, $\sigma_{\log P} \approx 1.3$ \\
\hline
\multirow{2}{*}{$0.7\,M_\odot \lesssim M_\star \lesssim 1.3\,M_\odot$} & $MF = 44\pm2\,\%$ & \multirow{2}{*}{$\gamma = 0.3 \pm 0.1$} & Unimodal (log-normal) \\
 & $CF = 62\pm3\,\%$ & & $\overline{a} \approx 45$\,AU, $\sigma_{\log P} \approx 2.3$ \\
\hline
\multirow{2}{*}{$1.5\,M_\odot \lesssim M_\star \lesssim 5\,M_\odot$} & $MF \geq 50\%$ & \multirow{2}{*}{$\gamma = -0.5 \pm 0.2$} & Bimodal \\
 & $CF = 100\pm10\,\%$ & & $\overline{P} \approx 10$d \& $\overline{a} \approx 350$\,AU \\
\hline
\multirow{2}{*}{$8\,M_\odot \lesssim M_\star \lesssim 16\,M_\odot$} & $MF \geq 60\%$ & \multirow{2}{*}{\dots} & \multirow{2}{*}{\dots} \\
 & $CF = 100\pm20\,\%$ & & \\
\hline
\multirow{2}{*}{$M_\star \gtrsim 16\,M_\odot$} & $MF \geq 80\%$ & $\gamma^{P \leq 3000\,{\rm d}} = -0.1 \pm 0.6$ & Peak + power law \\
 & $CF = 130\pm20\,\%$ & $\gamma^{a \geq 100\,{\rm AU}} = -0.5 \pm 0.1$ & $\overline{P} \approx 5$d \& \"Opik (?)\\
\botrule
\end{tabular}
\end{table}

\subsubsection{Multiplicity frequency}

\begin{figure}
\includegraphics[width=\textwidth]{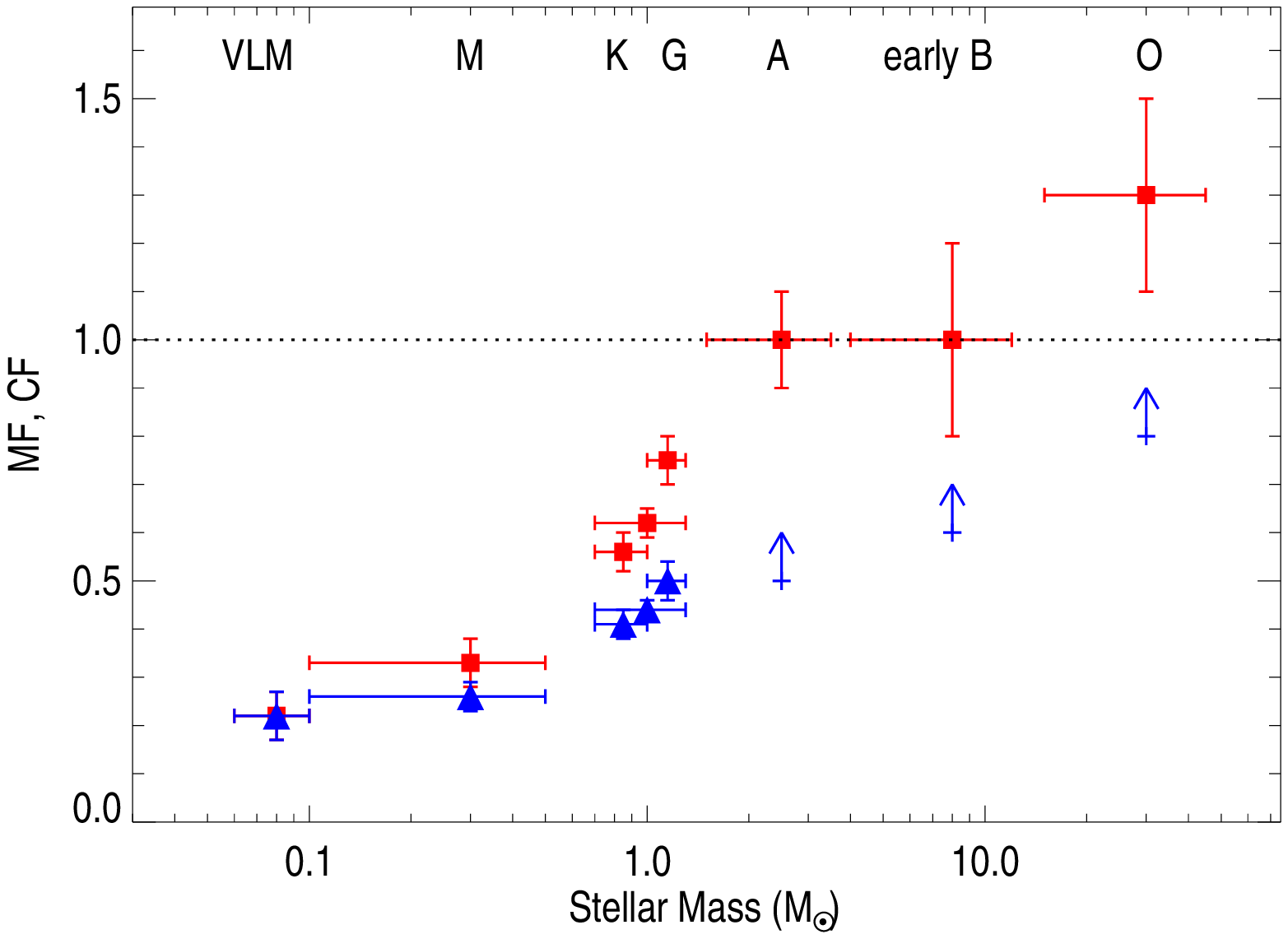}
\caption{Dependency of $CF$ (red squares) and $MF$ (blue triangles) with primary mass for MS stars and field VLM objects. The horizontal errorbars represent the approximate mass range for each population. For B and O stars, only companions down to $q \approx 0.1$ are included. The frequencies plotted here are the best-estimate numbers from Sections\, 3.1--3.5, also reported in Table\,\ref{tab:multMS}.}
\label{fig:CF_mass}
\end{figure}

As has long been known, the overall multiplicity frequency of MS stars is a steep, monotonic function of stellar mass (see Figure\,\ref{fig:CF_mass}). Given the limited number of points available and the fact that companions to high-mass stars with $q \lesssim 0.1$ are not accounted for, we refrain from proposing a functional form that reproduces the intrinsic dependency of $CF(M_\star)$. The dependency of $MF(M_\star)$ is qualitatively similar to that of $CF(M_\star)$ except for a shallower slope resulting primarily from the fact that $MF$ has an intrinsic upper bound at 100\%.

The frequency and order of multiple systems also are steep functions of stellar mass. For instance, all 11 sextuple systems in the Multiple Star Catalog are A-type stars \citep{eggleton08}, while a single such candidate system is known among nearby solar-type stars \citep{raghavan10}, and the highest-order multiple system containing only low-mass stars currently known has 4 components \citep{reid05}. While it appears that the ratio of binary to higher-order systems does not vary much for objects with $M_\star \leq 1.5\,M_\odot$, it probably rises significantly toward high-mass stars.

\subsubsection{Orbital period distribution}

The distribution of orbital periods is unimodal for solar-type and lower-mass stars, but both the median separation and width of the distribution decrease sharply with decreasing stellar mass. As a result, the frequency of companions in the 1--10\,AU range (10--15\%) does not vary significantly with stellar mass for $M_\star \leq 1.5\,M_\odot$, including substellar objects. The much lower overall multiplicity of field BDs traces a marked deficit of wide companions as opposed to a uniform depletion over all separations. 

Intermediate- and high-mass stars have more complex distributions of orbital periods. A strong peak at the shortest periods ($\log P \approx 0$--1) is found in both populations with an amplitude that increases with stellar mass. VBs show a peak for intermediate-mass stars, while the situation is less clear for high-mass stars, for which a shallow power-law is currently preferred. Interestingly, the frequency of companions in the 1--10\,AU range among intermediate-mass appears in reasonable agreement with that observed among solar-type stars. 

\subsubsection{Mass ratio distribution}
\label{subsec:trend_q}

\begin{figure}
\includegraphics[width=\textwidth]{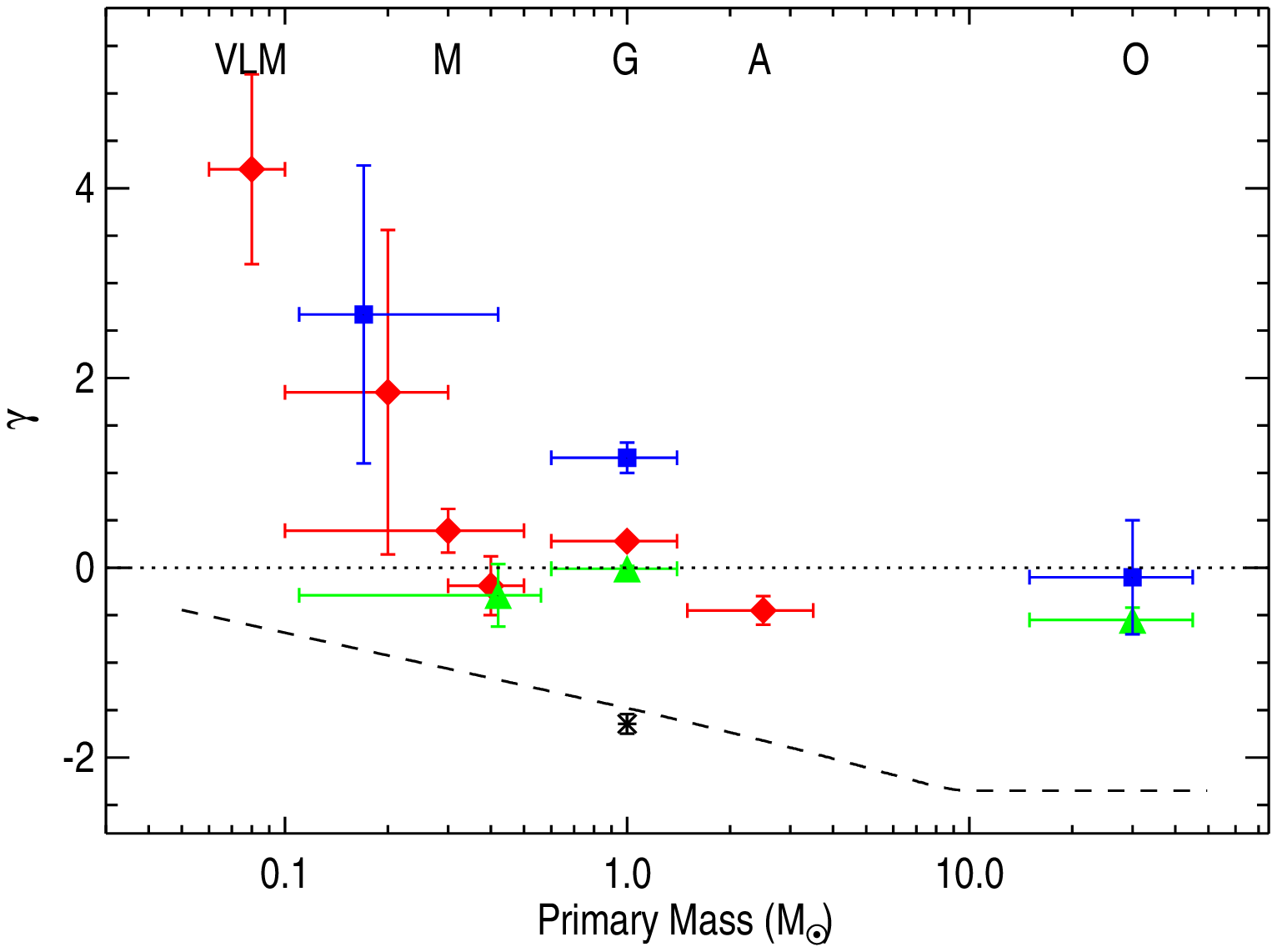}
\caption{Power law index fitted to the observed distribution of mass ratios for multiple systems as a function of primary mass. Red diamonds represent fits to the overall population of multiple systems within a certain range of primary masses, whereas blue squares and green triangles represent fits to the subsets of ``tight'' ($P \le \overline{P}$) and ``wide'' ($P \geq \overline{P}$) binaries, respectively. Horizontal errorbars represent the mass range for each subsample. The dashed curve represents the index that would be derived if the companions followed the single stars IMF of \cite{chabrier03} and a simple power law would be fit to the resulting companion mass distribution over the mass ratio range $0.1 \leq q \leq 1$. The asterisk marks the derived power law index using the single stars IMF of \cite{bochanski10}. The power law indices plotted here are the best-estimate numbers from Sections\, 3.1--3.5, also reported in Table\,\ref{tab:multMS}.}
\label{fig:alphaq_mass}
\end{figure}

Although a simple power law representation is imperfect for most samples, this formalism offers the most straightforward criterion to compare multiple systems of various masses. As shown in Figure\,\ref{fig:alphaq_mass}, the observed distribution of mass ratios is close to a flat distribution ($|\gamma| \lesssim 0.5$) down to $q \sim 0.1$ for all masses $M_\star \gtrsim 0.3\,M_\odot$, extending to high-mass stars the conclusions of \cite{reggiani11}. Below this limit, the mass ratio distribution becomes increasingly skewed towards high-$q$ systems. Furthermore, shorter-period systems among solar-type and low-mass stars have a steeper mass ratio distribution than wider systems.

Many past studies have favorably compared the mass distribution of companions to that expected from random pairing from the IMF for field objects. In this situation, and leaving aside the ``smearing'' induced by broad ranges of primary masses \citep{tout91}, $\gamma$ should increase from -2.3 (Salpeter's slope) at the high-mass end to $\sim 0$ around or below the substellar limit, which does not match observations at any primary mass. Although a proper comparison between observations and theory/simulations should be conducted on the raw $f(q)$ distributions rather than on estimated power law indices, the hypothesis of random pairing from the IMF is robustly excluded by observations (see Figure\,\ref{fig:alphaq_mass}). On the other hand, the rarity of substellar companions to low-mass stars suggests an alternative model in which the mass ratio distribution is flat between $q=1$ and a minimum companion mass, $M_{\rm sec}^{\rm min}$, that is independent of the primary mass (at least up to $M_\star \sim 1.5\,M_\odot$) and lies around or somewhat below the substellar limit. However, given current statistical uncertainties, it is also possible that there is no such minimum companion mass, or that it varies with primary mass.

\subsubsection{Eccentricity distribution}
\label{subsec:ecc}

Early studies of the eccentricity distribution in stellar binaries were hampered by severe selection biases \citep[e.g.,][]{aitken32, heintz69}. More modern systematic surveys have now yielded a clear and uniform picture \citep[e.g.,][]{abt90, raghavan10, derosa12, sana12, konopacky10, dupuy11, kiminki12b}. 

\begin{figure}
\includegraphics[width=\textwidth]{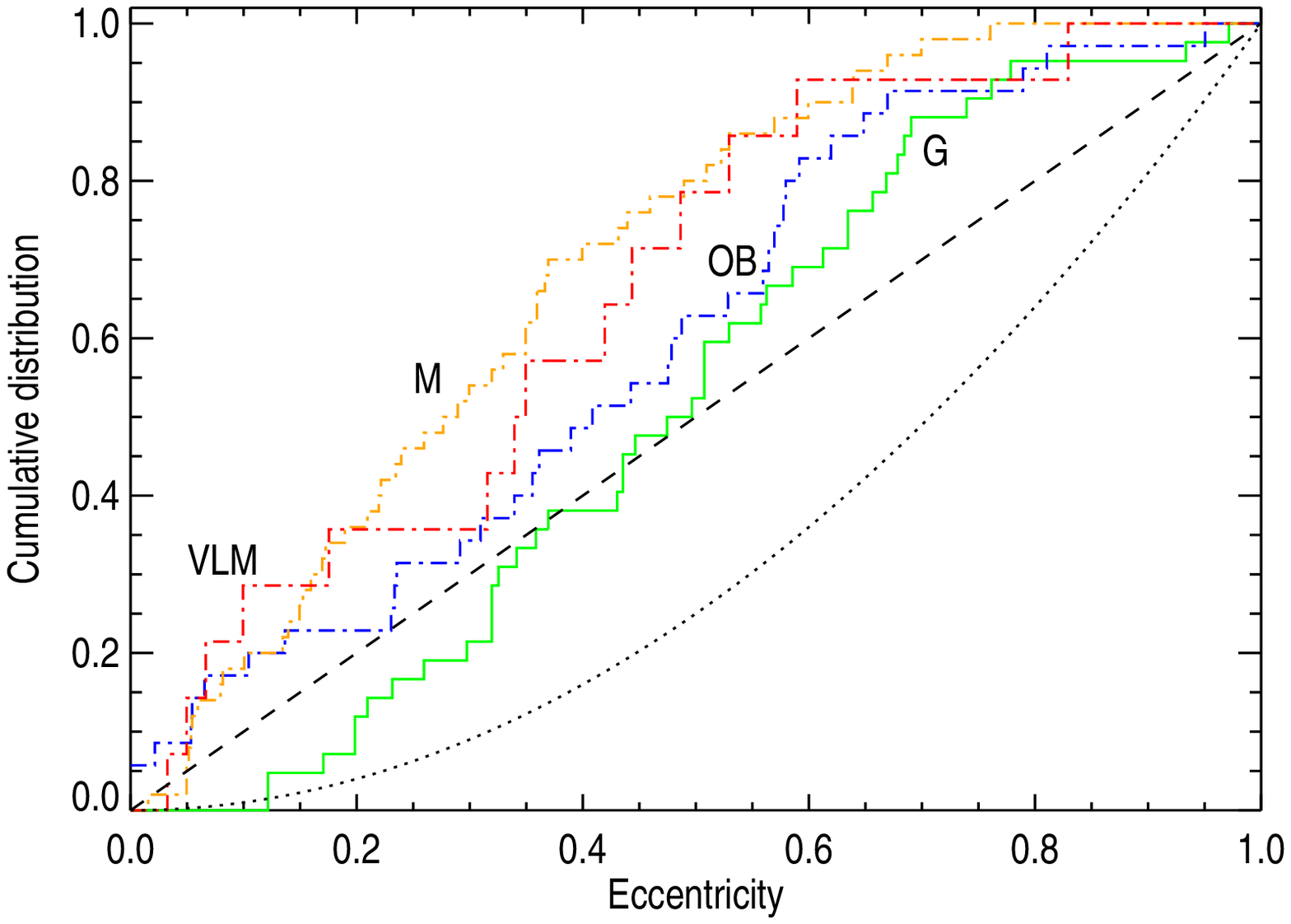}
\caption{Cumulative distribution of eccentricities for systems with $2 \le \log P \le 4$ for field multiple systems among solar-type stars \citep[green curve;][]{raghavan10}, low-mass stars \citep[orange curve; from the SB9 catalog, ][]{pourbaix04}, VLM stars and BDs \citep[red curve;][]{dupuy11}, high-mass stars \citep[blue curve; from the SB9 catalog and][]{abt05,sana12}. The dot-dashed curves indicate incomplete samples, for which the eccentricity distribution is potentially biased. The dashed and dotted curves represent the expected distribution for a flat and thermal distribution, $f(e)=2e$, respectively.}
\label{fig:ecc}
\end{figure}

The eccentricity distribution for field multiple systems shows remarkably little dependency on primary mass, extending the conclusions of \cite{abt05, abt06} across the entire range of primary masses. Beyond a period of $\sim 100$\,d, all populations show an essentially flat distribution\footnote{Although \cite{dm91} favored such a distribution, it clearly violates their bias-corrected distribution, which is flat beyond $e\approx 0.3$ (see their Fig.\ 6b).}. As illustrated in Figure\,\ref{fig:ecc}, a broad Gaussian distribution centered on $\overline{e} \approx 0.4$ is a possible alternative representation, since it produces somewhat less high-eccentricity systems \citep[e.g.,][]{stepinski01}. Discriminating between these two representations hinges on large, unbiased samples, which are still missing for most stellar masses. In any case, the observed distributions are all inconsistent with the so-called ``thermal distribution'', even though the latter is frequently assumed in numerical simulations. 

Almost all short-period systems have circular orbits as a result of tidal dissipation. The observed circularization periods, which range from $\lesssim 1$\,d to $\approx 20$\,d, is a function of the primary mass and a monotonic function of stellar age, from open clusters members to Population\,II stars. A handful of exceptions with eccentric short-period binaries are known among solar-type binaries. All of them are either members of higher-order systems that may be undergoing Kozai cycles, and/or relatively young so that they have not fully circularized yet \citep{mathieu88}.

Between these two regimes, the eccentricity distribution has a well-defined rising upper envelope in the $P$-$e$ plane. The mild correlation between mean eccentricity and orbital period, first proposed by \cite{finsen36} and borne out by recent surveys, is mostly due to the dearth of systems with $e \gtrsim 0.6$ at orbital periods 10--300\,day. Intriguingly, extrasolar planets populate the $P - e$ plane in a very similar manner as stellar multiple systems \citep{udry07}. 

\subsection{Multiplicity as a function of age}
\label{subsec:trend_time}

\indent \indent We now turn to the temporal evolution of multiplicity for solar-type and lower-mass stars, which we can probe for ages ranging from $\lesssim 0.1$\,Myr for the most embedded protostars to $\gtrsim 10$\,Gyr for Population\,II field stars. Higher-mass stars are only probed over a much narrower age range and do not allow for as thorough an analysis.

\subsubsection{Multiplicity frequency}

No population other than nearby field stars has been sufficiently studied to draw a complete picture of multiplicity, so this analysis can only be achieved for certain categories of companions. For SBs, no significant difference is observed between TTS, open cluster members, Population\,I and II field stars, although this conclusion is potentially weakened by incompleteness and biases and deserves further attention. For the remainder of this section, ``multiplicity frequency'' specifically refers to the frequency of visual companions. 

Even among VBs, only a narrow range of separations ($\sim 50$--500\,AU) has been studied across all populations, due to the diversity of distances involved and observational techniques employed. This limited range is insufficient to draw statistically robust conclusions, given typical sample sizes. Instead, we proceed by estimating for each population the companion frequency per decade of projected separation using the widest possible separation range within the approximate 10--3000\,AU range. This implicitly assume that all population of multiple systems follow \"Opik's law. Although an imperfect fit to the overall observed distribution, it is a reasonable approximation in the separation range considered here (which is near the peak of most proposed log-normal separation distributions).

\begin{figure}
\includegraphics[width=\textwidth]{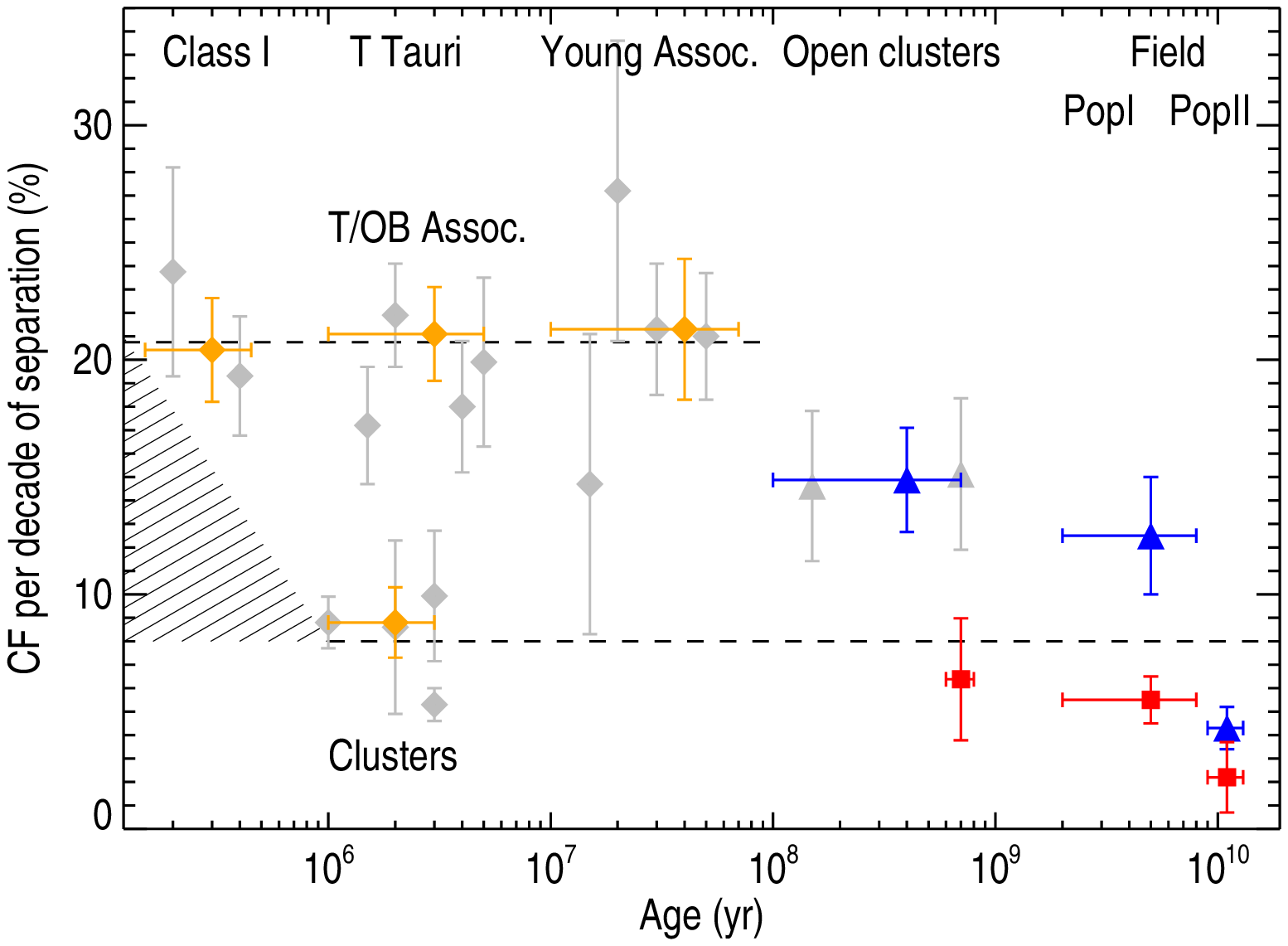}
\caption{Dependency of the frequency of visual companions per decade of projected separation with age for solar-type ($\sim 0.7$--1.5\,$M_\odot$, blue triangles), low-mass ($\sim 0.1$--0.5\,$M_\odot$, red squares) and overall populations of young stars ($\sim 0.1$--2\,$M_\odot$, orange diamonds). Results from individual surveys are indicated as gray symbols and include \cite{duchene07} and \cite{connelley08} for Class\,I sources, \cite{ratzka05}, \cite{kraus08, kraus11}, and \cite{lafreniere08} for T/OB associations, \cite{reipurth07}, \cite{beck03}, \cite{duchene99} and \cite{luhman05} for clusters of TTS, \cite{brandeker03, brandeker06}, \cite{chauvin10} and \cite{mccarthy12} for nearby young associations, and \cite{bouvier97, bouvier01} for open clusters. The horizontal errorbars represent the approximate age range for each population while vertical errorbars compound statistical uncertainties and the observed dispersion between similar regions. The dashed lines represent constant frequencies describing qualitatively the predicted behavior of low-density associations (top line) and dense clusters (bottom line). The hashed region represent the range of possible behaviors for stellar clusters in the embedded phases.}
\label{fig:CF_time}
\end{figure}

The resulting frequency of visual companions is shown in Figure\,\ref{fig:CF_time}, where we have elected to focus on the largest samples studied with uniform and complete techniques whenever possible. In addition, in cases where several surveys address the same ``population'' (e.g., several open clusters), we average the resulting multiplicity frequency and consider the dispersion of individual results as an additional source of uncertainty to account for some inherent diversity. Among TTS, we group separately dense clusters from loose associations, since their multiplicity frequency is markedly different. 

Comparisons between populations of young stars and that of field stars is complicated by the broad range of stellar masses probed in the early stages of stellar evolution, where surveys typically include objects of all masses from the substellar limit up to $\sim 2\,M_\odot$. In the past, it has been customary to compare most PMS multiplicity surveys to the well-studied population of solar-type field stars. However, a fairer comparison should instead be performed with a weighted average of solar-type and low-mass stars that takes into account possible selection biases in the young stellar populations. 

Figure\,\ref{fig:CF_time} highlights two parallel trends. On the one hand, stellar multiplicity does not vary on timescales $\lesssim 50$\,Myr in low-density environment (star-forming regions and young associations). On the other hand, there is no significant difference in multiplicity frequency between young and open clusters on one hand, and Population\,I field star on the other hand. It remains undetermined whether the dichotomy of multiplicity properties at early ages necessarily implies that the initial multiplicity properties are environment-dependent. This is discussed in more detail in Section\,\ref{sec:implications}. Finally, the marked multiplicity deficit among Population\,II stars, which is limited to wide binaries, suggests either a metallicity dependence in binary formation or evolution on a timescale of several Gyr. 

\subsubsection{Other multiplicity properties}

Beyond the evolution of the multiplicity frequency, most of the trends identified among PMS multiple systems closely match those observed among open cluster members and field stars. This includes the dependence of the mass ratio distribution on both stellar mass and separation, and the dependence of the maximum separation with stellar mass \citep[e.g.,][]{kraus11}. Possibly the most discrepant property when comparing young and mature multiple systems is the orbital period distribution: among young stellar objects it appears closer to \"Opik's law than among field stars, especially for solar-type systems \cite[e.g.,][]{connelley08, kraus11}. However, this only applies for young multiple systems in loose associations. Given the marked deficit of wide companions in young clusters like the ONC, averaging the orbital period distribution over all star-forming regions (weighted by their relative star formation rates) would result in a distribution that is much more similar to that observed among field stars. Not surprisingly, it thus appears as if multiplicity properties of field stars can be explained from those of TTS without much evolution. 

\subsection{``Twin'' binaries}
\label{subsec:twins}

\indent \indent The existence of a distinct population of ``twin'' binary systems has been repeatedly advocated and is still subject to intense debate, in part because of the loose definition of this phenomenon. Broadly speaking, twin binaries are near equal-mass, short-period binaries which are over-represented over the underlying population of binaries \citep[e.g.,][]{lucy79}. 

An excess of systems with orbital period $\leq$43\,d (median period $\sim 7$\,d) and $q \geq 0.98$ among 0.5--2\,$M_\odot$ binaries has been detected \citep{lucy06, simon09} although this subpopulation represents only 2--3\% of all SBs. This low frequency explains why it is not readily apparent in volume-limited surveys \citep{raghavan10}. It is probable that this excess results from accretion in the early stellar evolution stages that acts to equalize the components' masses. 

\cite{pinsonneault06} found an excess of systems with $q \ge 0.87$ among eclipsing binaries among 7--27\,$M_\odot$ stars in the Small Magellanic Cloud. However, \cite{lucy06} suggested that this trend could be explained in large part by selection biases and subsequent studies of Galactic high-mass stars do not find evidence for a separate population of twin systems (see Section\,\ref{subsec:twins}). Still, it is plausible that twin binaries may be a feature associated with low metallicity, since high-mass binaries in the Large Magellanic Cloud show a marginally steeper mass ratio distribution than their Galactic counterpart \citep{sana12b}.

\subsection{``Extreme'' mass ratio systems}
\label{subsec:extreme}

\begin{figure}
\includegraphics[width=\textwidth]{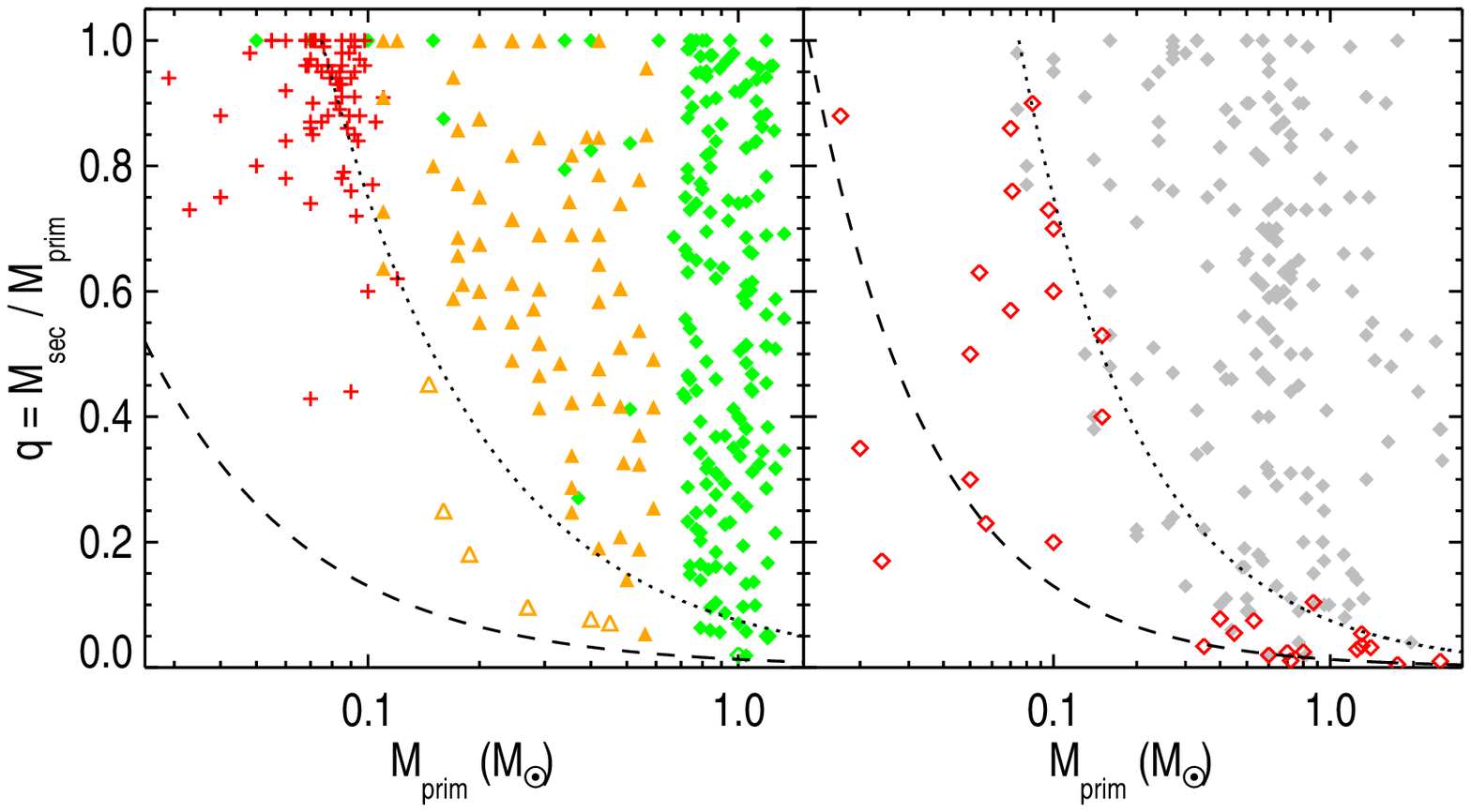}
\caption{{\it Left:} Distribution of mass ratios as function of mass for nearby field objects with $M_\star \le 1.5\,M_\odot$. The dotted and dot-dashed curves indicate constant companions masses at the substellar and planetary regime limits (0.075 and 0.013\,$M_\odot$), respectively. Data for solar-type stars (green diamonds), low-mass stars (orange triangles) and VLM objects (red plus signs) are from \cite{raghavan10}, the RECONS survey and \cite{janson12}, and the VLM binary database, respectively. BD companions to low-mass stars identified outside of large-scale surveys are shown as open orange triangles. The few systems with $M_{\rm prim} \le 0.5\,M_\odot$ from \cite{raghavan10} correspond to lower mass subsystems in hierarchical multiple systems. {\it Right:} Similar plot for star-forming regions and young associations. Shown as gray diamonds in this plot are companions to Class\,II/III objects in Taurus, Upper Scorpius and Chamaeleon\,I, using the large surveys from \cite{lafreniere08, kraus08, kraus11, kraus12}. In addition, red open diamonds represent a number of objects discovered via small-scale surveys or pointed observations with a particular emphasis on VLM primaries and/or companions.}
\label{fig:q_mprim}
\end{figure}

\indent \indent The interest in ``extreme'' mass ratio systems (broadly defined as $q \lesssim 0.1$) has rapidly grown, both in the context of the search for planetary-mass companions and because of the difficulty for binary formation model to produce them \cite[e.g.,][]{bate12}. Recent surveys have largely overcome the contrast limitations of earlier searches, providing new insights into the formation of VLM companions. Figure\,\ref{fig:q_mprim}, which is based on large-scale surveys of field and young stars, illustrates the marked deficit of low-$q$ systems, especially among (very) low-mass objects. 

The extremely low frequency of BD companions among solar-type SBs \citep{marcy00, grether06} is inconsistent with extrapolations of both the binary mass function and the planetary mass function \citep[e.g.,][]{howard10, sahlmann11, reffert11}. This suggests a sharp discontinuity in companion masses around the hydrogen burning limit \citep{thies08}, although this BD ``desert'' is not completely arid and warrants further analysis. 

Among visual companions, the stringent limits on the presence of substellar companions obtained a decade ago \citep[$\le$1\% ; e.g.,][]{oppenheimer01, hinz02, mccarthy04} applied to samples largely dominated by low-mass stars. More recently, deeper surveys focusing on solar-type stars found potentially higher frequencies, although the very small number of detections limits the robustness of this result \citep{carson06, metchev09, leconte10, tanner10}. It remains unclear whether these substellar companions are a simple extension of the population of companions below 0.075\,$M_\odot$, as proposed by \cite{metchev09}, or whether a distinct trough in companion mass is present at large separations as for short-period systems. One approach to test whether there is a demarcating line between two distinct populations of visual companions consists in studying the frequency of (very) low-mass stellar companions to intermediate- and high-mass stars, which probe similar mass ratios. In this context, we note the existence of a handful of planetary-mass objects around intermediate-mass stars \citep{marois08, kalas08} but an apparent dearth of companions around the substellar limit \citep{janson11, vigan12}. Upcoming surveys for wide planetary mass companions to nearby stars using ``extreme'' adaptive optics instruments will help address this issue in a statistically sound manner.

Multiple surveys have addressed the existence of the BD desert among visual companions in both star-forming regions and nearby associations \citep[e.g.,][]{kouwenhoven07a, tanner07, evans12}. These survey generally found that BD companions are rare but not absent \citep[e.g.,][]{lafreniere11, janson12b, nielsen12}. Finally, high-resolution imaging surveys also have discovered a small number of planetary-mass companions ($\sim 4$--20 $M_{Jup}$) at distances of $\gtrsim100$\,AU from young stars in nearby star-forming regions \citep{chauvin04, neuhauser05, luhman06, lafreniere08b, schmidt08, ireland11}. Unfortunately, most of these discoveries were serendipitous, precluding a robust statistical analysis of their frequency. 

\subsection{Extremely wide systems}
\label{subsec:wide}

\indent \indent Observationally, the correlation between the median separation and primary mass observed for stars with $M_\star \lesssim 2\,M_\odot$ is accompanied by a similar trend for the maximum binary separation $a_{\rm max}$ \citep[e.g.,][]{reid01}. Although different functional forms and an unconfirmed discontinuity at $\sim 0.2\,M_\odot$ have been proposed \citep{reid01, burgasser03, close03}, this upper limit on binary separation appears to hold for both field populations and PMS systems \citep[e.g.,][]{konopacky07, kraus09, faherty11}. Furthermore, several authors used energetic arguments to propose that the maximum binary separation is actually set as a minimum binding energy \citep[e.g.][]{close03}. It remains undetermined which quantity is the most physically relevant.

A key to understanding the physical meaning of the $a_{\rm max} - M_{\rm tot}$ correlation lies in ``unusually wide'' systems, which violate the general trend. While the vast majority of field binaries fall inside the maximum separation limit, numerous studies have successfully searched for systems well beyond that limit. For solar-type and intermediate-mass stars stars, systems with separation out to several pc have been identified \citep[e.g.,][]{caballero09, caballero10, shaya11}. Among field low-mass stars, for which the term ``unusually wide'' was first coined, systems at separations $\geq10^3$ AU have been found \citep[e.g.,][]{artigau07, caballero07, radigan09, dhital10, faherty11}. A number of unusually wide low-mass systems have also been found in nearby star-forming regions \citep{luhman09, kraus09, bejar08, allers09}. Around the substellar limit, although a marked paucity of companions is observed beyond $\gtrsim$25\,AU, a small number of wide pairs have been identified, particularly among young objects \citep[e.g.,][]{chauvin04, luhman06, kraus07b, bejar08}. In all cases, these ``unusually wide'' systems have separations $a \gtrsim 10 a_{\rm max}$. Even though follow-up observations of some of these wide systems indicate that they are frequently part of hierarchical multiple systems \citep{law10}, their binding energy is still well below the empirical limit discussed above. 

The existence of these unusually wide systems, coupled with the fact that the empirical limit on binary separation has been repeatedly revised upwards over the last decade, suggests that there is no absolute upper bound on binary separation, but rather a gradual rarity beyond a certain separation.

\section{IMPLICATIONS FOR BINARY FORMATION AND EVOLUTION}
\label{sec:implications}

\indent \indent 
Multiplicity is a smooth function of stellar mass, both on the MS and in the earlier phases of stellar evolution. This trend points to a similar formation process across a broad range of prestellar core masses. Core (``prompt'') fragmentation still appears as the leading candidate scenario to form multiple systems, although the extreme diversity of systems, most notably the large breadth of the orbital period distribution, may require at least two separate mechanisms \citep{tohline02}. 

Fragmentation appears to be mild, in that most cores only produce between 1 and 3 components. Since the frequency of quadruple and higher-order systems among field stars is low, most high-order systems would need to be destroyed early on if core fragmentation typically led to a larger number of stellar seeds. This destruction would result in a much lower overall multiplicity frequency among field stars, especially among low-mass stars, since many ejected components would be single stars. While this is supported by purely dynamical star formation simulations \citep[e.g.,][]{delgado04}, it is not borne out by observations \citep[see also][]{goodwin05}. Numerical simulations including the effects of magnetic field and/or radiative feedback appear to match more closely observed multiplicity properties since these processes severely hinder fragmentation \citep[e.g.,][]{hennebelle08, bate12}.

The statement that ``most stellar systems formed in the Galaxy are likely single and not binary'' \citep{lada06} is still undecided. While it is true that most field stars are single as a result of the preponderance of low-mass stars, it is contrary to observations in low-density star-forming regions. It is unclear whether these regions are representative of multiple star formation in all environments. Indeed, the universality of multiplicity properties among different birth environments is still in doubt. There are marked observational differences, most notably the lower multiplicity observed in clusters relative to loose associations of young stars. However, numerical simulations show that (internal and cluster) dynamical evolution is by and large over after 1\,Myr \citep[e.g.,][]{marks12}, suggesting that current observations do not probe multiplicity in an early enough evolutionary stage to be conclusive. Because frequent interactions in dense clusters can reduce the initial multiplicity frequency by a factor of 2--4, a picture in which multiplicity frequency is universally high at early times remains plausible \citep{kroupa11}. One critical test would be to determine the multiplicity properties of clustered protostars since Class\,I surveys have so far only targeted objects born in ``quiet'', low stellar density environments. Another possible test of this scenario consists in comparing the mass-dependency of multiplicity in young clusters and T/OB associations, since dynamical simulations predict that lower mass systems are more prone to disruption \citep{kroupa01}. 

Dynamical evolution can naturally explain both the ``low multiplicity'' and ``high multiplicity'' tracks observed in Figure\,\ref{fig:CF_time}, in which clusters and loose associations represent extreme cases of highly effective and ineffective binary disruption \citep{kroupa03}. Irrespective of whether initial multiplicity properties are universal, the existence of these two tracks has immediate implications on the origin of field stars. Indeed, it appears that most stars form in clusters, an argument already made in the past based on multiplicity \citep{patience02, parker09, marks11} as well as from star/cluster counts \citep{lada03}. A more thorough analysis that goes beyond the multiplicity frequency and also considers other parameters, such as the orbital period distribution or dependencies on the primary mass, is still needed.

The fact that the mass ratio distribution is essentially flat for most primary masses ($M_\star \gtrsim 0.3\,M_\odot$) but becomes increasingly steeper at lower masses (Section\,\ref{subsec:trend_q}) sets important constraints on both binary formation itself and the subsequent accretion through which final stellar masses are built. In particular, the suggestion that there may be a lower limit to companion masses, which remains unconfirmed with current observations, could be pointing to the smallest scale associated with fragmentation. In addition, the (small but growing) population of planetary-mass objects at large distances from their parent star (Section\,\ref{subsec:extreme}) poses a significant challenge to existing models of star and planet formation. Their orbital radii are too large for a ``planetary'' formation model \citep[i.e., from a protoplanetary disk;][]{pollack96, boss01}, but their masses are extremely low for a ``binary'' formation model \citep[falling near or below the opacity-limited minimum mass;][]{kratter10}. Nonetheless, it appears that one of these unlikely outcomes must occur. Determining whether these objects represent an extension of (sub)stellar companions to extremely low masses or form a separate population of objects will be addressed in the next few years. 

At short orbital periods, although the long-discussed BD desert has been repeatedly confirmed as tracing physically distinct formation mechanisms, the remarkable similitude between the eccentricity distributions of planetary and stellar-like companions (including massive BDs) is intriguing. One possible interpretation is that the final eccentricity in both planetary and multiple stellar systems is ultimately set by dynamical $N$-body interactions. If this is true, it remains to be understood why the eccentricity distribution of multiple systems does not follow the thermal distribution, however.

The observed correlation between primary mass and ``maximum'' binary separation is plausibly linked to the typical prestellar core size \citep[both on the order of $\sim 0.1$\,pc][]{bahcall81, lada08}. Similarly, the median binary separation, which also scales with primary mass, could be hinting at a preferred spatial scale for fragmentation. Alternatively, the physical origin of $a_{\rm max}$ could lie in the dynamical evolution of systems. Indeed, the widest systems have a very weak binding energy that makes them vulnerable to disruption by other nearby passing stars, either in their birth environment \citep{kroupa95, parker09} or in the Galactic field \citep{weinberg87, jiang10}. The latter secular evolution may be responsible for the lower multiplicity observed among Population\,II stars at large separations. A constant minimum binding energy can be expected in dynamically relaxed populations, such as virialized stellar clusters or field stars. However, even dynamically primordial environments like Taurus-Auriga and Upper Scorpius broadly follow the same $a_{\rm max}$ trend, casting doubt on this interpretation.

The physical origin of unusually wide systems (Section\,\ref{subsec:wide}) is not fully ascertained. They could form in a similar manner as other binaries from a single core. Since they would not survive dynamical interactions in dense clusters, one possible explanation for their extreme paucity in the field population is that ``unusually wide'' binaries form via a low-probability channel, and that only those binary systems born in unbound associations can remain bound until they reach the field \citep{kraus08, kraus09}. The main $M_{tot}-a_{max}$ relation would then indicate the typical constraint from protostellar core sizes or the maximum spatial scale for fragmentation. Alternatively, they may have formed from separate cores and have been ejected simultaneously from a cluster on sufficiently similar trajectories as to remain weakly bound \citep{kouwenhoven10, moeckel11}. The low probability of this scenario readily accounts for their rarity among field stars.

\section{SUMMARY POINTS}

\begin{itemize}
\item The frequency and key properties associated with stellar multiplicity vary smoothly with primary mass, suggesting a similar formation process across all core masses.
\item Multiplicity properties appear to be mostly set at the PMS phase. Subsequent evolution is modest at best, as expected from dynamical evolution models.
\item No single population of binary systems has a mass-ratio distribution that agrees with random pairing from the IMF. Also, no population has an eccentricity distribution that follows the ``thermal'' distribution, $f(e) = 2e$.
\item Observations of both PMS and field stars do not support the existence of a firm upper limit on binary separation, but an increasing rarefaction towards the largest scales (e.g., $\sim 1\,$pc for solar-type systems).
\item At short orbital periods ($\log P \lesssim 3$), there is a marked distinction between planetary- and stellar-type systems, presumably tracing different formation mechanisms.
\end{itemize}

\section{FUTURE ISSUES}

\begin{itemize}
\item Are the multiplicity properties resulting from the star formation process universal or do they depend on the native environment? The mass-dependency of multiplicity among PMS in different environments stars holds a critical clue.
\item Are there significant differences in the multiplicity properties of high-mass stars relative to lower-mass stars that would not follow a smooth mass dependency? A significant frequency of stellar companions with $q \lesssim 0.1$, or the confirmation of a separate but substantial population of ``twin'' binaries, would set high-mass stars apart.
\item What is the formation path and associated timescale for SBs? The frequency and properties of VBs appear to be set by the TTS phase, but is this also true for SBs?
\item What is the origin and fate of ``unusually wide'' systems?
\item Is the dearth of ``extreme mass ratio'' systems (known as the BD desert among tight solar-type and lower-mass binaries) independent of separation and primary mass? What are the physical processes responsible for this phenomenon?
\item Does multiplicity hinder or promote planet formation? How does it depend on binary separation, mass ratio, and eccentricity?
\end{itemize}

\section{LITERATURE CITED} 

\section{SIDEBAR}

{\em Note to editor: place alongside Section 2}

Short period binaries ($\log P \lesssim 4$) are best identified through radial velocity monitoring. Longer period (``visual'') systems are revealed by direct and increasingly higher-angular resolution imaging surveys using speckle interferometry, adaptive optics-aided imaging, the Hubble Space Telescope, aperture masking on monolithic telescopes, lunar occultation and long-baseline interferometry. Companions in the most challenging ``intermediate'' regime ($P \gtrsim 5$\,yr, $\rho \lesssim 0.05"$) can also be probed by astrometric monitoring, although it has been difficult to apply to large, well-defined stellar populations except for bright stars observed by {\sl Hipparcos}.

Multiplicity can also be probed indirectly. In uniform stellar populations, high mass ratio systems appear over-luminous, i.e., located up to 0.75\,mag above the sequence of single objects. The vertical spread in the HR diagram can thus be modeled statistically to constrain the multiplicity frequency and the mass ratio distribution. Alternatively, individual objects whose spectrum contains two distinct sets of photospheric features, revealing the presence of a companion of much different effective temperature, are known as ``spectral'' binaries. These are easiest to detect in cases where the $M_\star - T_{\rm eff}$ relationship is steep but the $M_\star - L_\star$ is shallow, most notably among VLM stars and BDs.

\section{Acronyms and Definition}

\begin{itemize}
\item $a$: semi-major axis (expressed in AU unless otherwise stated)
\item BD: brown dwarf
\item $CF$: companion frequency (average number of companion per target in a population)
\item IMF: initial mass function
\item $MF$: multiplicity frequency (fraction of multiple systems in a population)
\item MS: Main Sequence
\item ONC: Orion Nebula Cluster
\item $P$: orbital period (expressed in days unless otherwise stated)
\item PMS: Pre-Main Sequence
\item SB: spectroscopic binary (broadly speaking, $\log P \lesssim 3$)
\item TTS: T\,Tauri stars
\item VB: visual binary (broadly speaking, $\rho \gtrsim 0.05"$)
\item VLM: very low-mass
\item $\alpha$: index of the power-law fit to the orbital period distribution
\item $\gamma$: index of the power-law fit to the mass ratio distribution
\item $\rho$: angular separation of a visual binary (expressed in arcsec unless otherwise stated)
\end{itemize}

\end{document}